\documentclass[twocolumn,10pt]{article}

\usepackage[utf8]{inputenc}
\usepackage[T1]{fontenc}
\usepackage{amsmath,amsfonts,amssymb}
\usepackage{graphicx}
\usepackage{xcolor}
\usepackage{hyperref}
\usepackage{cite}
\usepackage{abstract}
\usepackage{titlesec}
\usepackage{geometry}
\usepackage{fancyhdr}
\usepackage{authblk}
\usepackage{booktabs}
\usepackage{multirow}
\usepackage{array}
\usepackage{colortbl}
\usepackage[table]{xcolor}
\usepackage{amsmath}
\usepackage{tabularx}

\hypersetup{
    colorlinks=true,
    linkcolor=blue,
    filecolor=magenta,      
    urlcolor=cyan,
    citecolor=blue,
}
\definecolor{lightgray}{gray}{0.95}
\definecolor{lightblue}{HTML}{E6F0FA}

\renewcommand{\arraystretch}{1.05}

\author[1]{Vipin Rathi}
\author[2]{Lakshya Chopra}
\author[2]{Madhav Agarwal}
\author[2]{Nitin Rajput}
\author[2]{Kriish Sharma\\}
\author[2]{Sushant Mundepi}
\author[2]{Shivam Gangwar}
\author[2]{Rudraksh Rawal}
\author[2]{Jishan}

\affil[1]{Ramanujan College, University of Delhi, New Delhi, India}
\affil[2]{coRAN Labs Private Limited, New Delhi, India}

\titleformat{\section}{\large\bfseries}{\thesection}{1em}{}
\titleformat{\subsection}{\normalsize\bfseries}{\thesubsection}{1em}{}
\titleformat{\subsubsection}{\small\bfseries}{\thesubsubsection}{1em}{}

\title{\vspace{-1cm}\textbf{Q-RAN: Quantum-Resilient O-RAN Architecture}}

\date{}

\DeclareUnicodeCharacter{202F}{\,}
\begin{document}

\maketitle

\definecolor{linkblue}{HTML}{1A4D8F}
\begin{abstract}
\textbf{The telecommunications industry faces a dual transformation: the architectural shift toward Open Radio Access Networks (O-RAN) and the emerging threat from quantum computing. O-RAN's disaggregated, multi-vendor architecture creates a larger attack surface vulnerable to cryptanalytically relevant quantum computers (CRQCs) that will break current public-key cryptography. The Harvest Now, Decrypt Later (HNDL) attack strategy makes this threat immediate, as adversaries can intercept encrypted data today for future decryption. This paper presents Q-RAN, a comprehensive quantum-resistant security framework for O-RAN networks using NIST-standardized Post-Quantum Cryptography (PQC). We detail the implementation of ML-KEM (FIPS 203) and ML-DSA (FIPS 204), integrated with Quantum Random Number Generators (QRNG) for cryptographic entropy. The solution deploys PQ-IPsec, PQ-DTLS, and PQ-mTLS protocols across all O-RAN interfaces, anchored by a centralized Post-Quantum Certificate Authority (PQ-CA) within the SMO framework. This work provides a complete roadmap for securing disaggregated O-RAN ecosystems against quantum adversaries.}

\textbf{\\Keywords---\textcolor{linkblue}{Post-Quantum Cryptography (PQC), 5G RAN (5G-RAN), O-RAN Alliance, Quantum-Safe Networks (QSN), Quantum-Resilient 5G RAN, Hybrid PQC (HPQC), Quantum-Safe Authentication, Post-Quantum Migration, Post-Quantum TLS (PQ-TLS), Post-Quantum IPsec (PQ-IPsec), Hardware-Accelerated PQC, Post-Quantum PKI Infrastructure, Quantum-Safe Telecom, Post-Quantum Security for B5G/6G, QORE, xFAPI}}

\end{abstract}

\section{Introduction}

The global telecommunications industry is undergoing a major shift. Open Radio Access Networks (O-RAN) mark a significant departure from traditional, single-vendor base station designs. By breaking these systems into modular parts connected through open, standardized interfaces, O-RAN introduces greater flexibility and innovation. Driven by the O-RAN ALLIANCE, this approach separates the gNodeB into three key components — the Radio Unit (RU), Distributed Unit (DU), and Centralized Unit (CU) — and supports multi-vendor deployments. It also enables advanced network automation through the RAN Intelligent Controller (RIC).

As the industry advances toward an open RAN architecture, it faces a growing threat: quantum computing. Cryptographically Relevant Quantum Computers (CRQCs) have the potential to break widely used public-key cryptosystems such as RSA and Elliptic Curve Cryptography (ECC) \cite{rfc_ed448}. Notably, Shor’s algorithm \cite{shor} can efficiently factor large integers and compute discrete logarithms, making it capable of defeating these asymmetric cryptography schemes. This poses a serious risk to authentication, key exchange, and digital signatures across the entire network infrastructure.

This threat is immediate due to the Harvest Now, Decrypt Later strategy, where adversaries capture encrypted data today for decryption once quantum computers become available. For telecommunications networks with sensitive data and operational lifecycles spanning decades, this creates urgent vulnerability. O-RAN's open, disaggregated architecture inherently expands the attack surface, making quantum-resistant security not optional but essential.

The solution lies in Post-Quantum Cryptography (PQC), a family of algorithms designed to resist both classical and quantum attacks. In August 2024, the U.S. National Institute of Standards and Technology (NIST) finalized the first PQC standards \cite{nist_first_pq}, providing a clear path forward. This paper presents Q-RAN, a comprehensive framework for migrating O-RAN networks to quantum-resistant security, addressing every interface and component with standardized PQC algorithms and crypto-agile implementation strategies.

\subsection{Contributions}

This paper makes the following contributions:

\begin{itemize}
    \item A comprehensive analysis of quantum threats to O-RAN architecture and identification of vulnerable interfaces and protocols
    \item Detailed implementation specifications for PQ-IPsec, PQ-DTLS, and PQ-mTLS using NIST-standardized algorithms (ML-KEM-768, ML-DSA-65)
    % \item Integration of Quantum Random Number Generators (QRNG) providing true and unpredictable randomness to cryptographic operations
    % \item Centralized Post-Quantum Certificate Authority (PQ-CA) within the SMO framework for unified trust management across the entire RAN.
    % \item Practical deployment and implementation ideas using open-source libraries such as strongSwan and OpenSSL with the OQS provider, including configuration details and integration into O-RAN software stacks.
    % \item Step-by-step roadmap for migrating from classical cryptography to a fully quantum-resistant O-RAN security setup, highlighting key challenges and future scope.
    \begin{itemize}
    \item Integration of Quantum Random Number Generators (QRNGs) to ensure cryptographic operations use truly random, non-deterministic entropy sources.
    \item A Post-Quantum Certificate Authority (PQ-CA) placed within the SMO framework, serving as the central trust point for all RAN components.
    \item Practical deployment using open-source tools like strongSwan and OpenSSL (with the OQS provider), along with configuration examples drawn from real O-RAN stacks.
    \item Complete roadmap outlining how classical security can realistically evolve into a quantum-safe O-RAN setup, with the main hurdles and open problems clearly stated.
    \end{itemize}
\end{itemize}

\subsection{Paper Organization}

% The paper starts with essential background on O-RAN architecture, security mechanisms, and post-quantum cryptography fundamentals. It then outlines the strategy for migrating classical O-RAN systems to quantum-resistant security including integration and usage challenges. Next, our solution, Q-RAN, is described, including protocol specifications and integration with relevant libraries. Following this, deployment considerations, standardization requirements, and directions for future research are discussed. The paper concludes with a summary of key findings and insights gathered from the upgradation.

This paper starts by introducing the key elements of the O-RAN architecture, including its modular structure and built-in security mechanisms. It also offers a brief overview of post-quantum cryptography (PQC) to set the stage for the challenges ahead. From there, we explain why current O-RAN systems need to be adapted to resist quantum-era threats and walk through the steps involved in upgrading the existing 5G security framework. This includes proposed protocols, implementation enhancements, and a look at the practical hurdles of moving to PQC. We then present Q-RAN, our end-to-end tested solution that puts these ideas into practice. Finally, we share the main insights from our work and outline possible directions for future development.

\section{Background}

This section takes a closer look at the O-RAN architecture, focusing on its modular design and security features that are already in place. We then introduce the basics of post-quantum cryptography and talk about how current security protocols need to evolve to stay protected against future quantum threats.

\subsection{O-RAN Architecture}

Open RAN disaggregates traditional base station functions into interoperable components \cite{oran_arch}. While this encourages innovation, it also often leads to a larger attack surface. Because of its open design, Open RAN needs stronger, quantum-resistant security mechanisms.

\subsubsection{Logical Components}

\textbf{O-RAN Centralized Unit (O-CU):} This unit handles higher-layer protocol functions and is split into the Control Plane (O-CU-CP) and User Plane (O-CU-UP). The O-CU-CP manages RRC (Radio Resource Control) signaling and mobility management, while the O-CU-UP processes user data through the PDCP (Packet Data Convergence Protocol) layer, including encryption.

\noindent\textbf{O-RAN Distributed Unit (O-DU):} The DU is responsible for handling real-time layers---RLC, MAC, and High-PHY layers, handling tasks such as error correction, retransmission, scheduling, and modulation mapping. It interfaces with the RU over the Open Fronthaul (O-FH) and communicates with the CU using the F1 Application Protocol (F1AP).

\noindent\textbf{O-RAN Radio Unit (O-RU):} The RU is responsible for Low-PHY processing and the transmission and reception of radio signals over the air interface. It handles tasks such as signal modulation and demodulation, analog-to-digital conversion, as well as MIMO processing and beamforming. All these operations must be performed within tight timing and computational constraints..

\subsubsection{Service Management and Orchestration (SMO)}

The SMO orchestrates the RAN's disaggregated components through a variety of management interfaces, delivering FCAPS capabilities via the O1 interface and handling orchestration tasks through O2. It also hosts the Non-RT RIC, where rApps enhanced with AI/ML are responsible for policy enforcement and optimizations. As a result of its central function, the SMO is well suited to function as both a Post-Quantum Certificate Authority (PQ-CA) and a Post-Quantum Authorization Server, serving as a unified trust anchor for the network—a quantum-secure counterpart to a traditional root of trust or Post-Quantum PKI (PQ-PKI).

\subsubsection{RAN Intelligent Controllers}

\textbf{Near-RT RIC:} Controls O-CU/O-DU behavior on 10ms–1s timescales through xApps executing AI-guided radio resource management.

\noindent\textbf{Non-RT RIC:} Operates above 1s timescales, hosting rApps for long-term optimization and model training. It interfaces with the Near-RT RIC through the A1 interface for policy delivery.

\subsection{O-RAN Interfaces and Security Requirements}

O-RAN defines multiple interfaces with differing latency and security demands:

\noindent\textbf{A1:} Connects Non-RT and Near-RT RICs using mTLS for authenticated policy exchange.  \\
\textbf{E2:} Links Near-RT RIC with O-CU/O-DU, secured via IPsec; compromise could enable resource manipulation. \\ 
\textbf{F1:} Connects O-CU and O-DU—F1-C (control) and F1-U (user)—protected by IPsec/DTLS.  \\
\textbf{O1:} Management interface between SMO and O-RAN nodes, secured via mTLS.  \\
\textbf{O2:} Connects SMO to O-Cloud for lifecycle management, using mTLS.  \\
\textbf{Open Fronthaul:} High-throughput O-RU $\leftrightarrow$ O-DU interface; employs mTLS on the M-Plane and IEEE 802.1X authentication, but faces performance constraints for PQC adoption.

\subsection{Classical Security Mechanisms in O-RAN}

Current O-RAN security \cite{oran_security} relies on classical cryptographic primitives, such as RSA, DH and ECC, which are soon to be broken by quantum computers.

\noindent\textbf{Public Key Infrastructure (PKI):} PKI is a system concerned with establishing trust in cryptographic communications over insecure public networks. It typically achieves this with the use of digital X.509 Certificates \cite{rfc_x509}, which are used to confirm an entity's true identity. The certificates comprise digital signatures, created using classical cryptosystems like RSA / ECC, making them vulnerable to quantum attacks that are capable of forging certificates and breaking authenticity.\\
\textbf{IPsec:} Internet Protocol Security is a suite of open standards for achieving confidentiality, integrity, and authenticity over insecure IP connections via the use of cryptographic techniques and PKI \cite{rfc_ipsec_esp}. The protocol employs a set of asymmetric and symmetric primitives (e.g., AES) to guarantee security. Symmetric AES remains secure with 256-bit keys, but IKEv2 key exchange (e.g., Diffie-Hellman) is quantum-vulnerable. IPsec is the backbone of O-RAN security, being frequently deployed in various interfaces (N2, N3, E2, Xn). \\
\textbf{DTLS:} Datagram Transport Layer Security is a higher layer protocol designed to protect datagram oriented transport protocols (e.g., UDP, SCTP) (RFC 9147)~\cite{rfc_9147} using cryptographic techniques. Vulnerable in key exchange and authentication phases.  \\
\textbf{mTLS:} Mutual TLS extends the TLS protocol for guaranteeing mutual authentication between client and server. mTLS is mandated for A1, O1, O2, and M-Plane; both certificate and key exchange stages are quantum-breakable. \\
\textbf{OAuth 2.0:} An authorization framework designed to provide delegated access to protected resources (RFC 6749) ~\cite{rfc_6749}. Bound to mTLS certificates (RFC 8705)~\cite{rfc_8705}, but vulnerable if certificates can be forged via quantum attacks.

\subsection{Post-Quantum Cryptography Fundamentals}

Post-Quantum Cryptography (PQC) employs algorithms secure against both classical and quantum attacks. NIST-standardized lattice-based schemes—ML-KEM and ML-DSA—are key to future-proofing O-RAN security.

\subsubsection{ML-KEM (FIPS 203)}

ML-KEM \cite{nist_mlkem} is an IND-CCA2-secure Key Encapsulation Mechanism (KEM) derived from the CRYSTALS-Kyber \cite{crystals_kyber}. It enables secure shared key establishment through the standard \textbf{KeyGen, Encaps, and Decaps} operations. It is built upon the \textbf{Module Learning with Errors (MLWE)} problem, a widely studied hardness assumption in lattice-based cryptography that is believed to be resistant to attacks from both classical and quantum adversaries.

ML-KEM guarantees cryptographic security by leveraging the difficulty of distinguishing structured noisy linear equations from uniform randomness---\textbf{Decision LWE (D-LWE)}. The scheme samples error and noise terms from a centered binomial distribution, which offers both efficiency and compactness while preserving the required level of entropy for post-quantum security.

Parameter sets: ML-KEM-512 (AES-128 equivalent), 768 (AES-192), 1024 (AES-256), summarized in Table~\ref{tab:mlkem-sizes}.  
\textit{ML-KEM-768} balances efficiency and security for telecom deployments.
\begin{table}[h!]
\centering
\footnotesize
\setlength{\tabcolsep}{4pt} % tighter columns
\renewcommand{\arraystretch}{1.2}
\begin{tabular}{lcccc}
\rowcolor{lightblue}
\textbf{Set} & \textbf{Pub (B)} & \textbf{Sec (B)} & \textbf{Ct (B)} & \textbf{SS (B)} \\
ML-KEM-512  & 800  & 1,632 & 768  & 32 \\
ML-KEM-768  & 1,184 & 2,400 & 1,088 & 32 \\
ML-KEM-1024 & 1,568 & 3,168 & 1,568 & 32 \\
\end{tabular}
\caption{ML-KEM parameter set sizes (bytes)}
\label{tab:mlkem-sizes}
\end{table}

\subsubsection{ML-DSA (FIPS 204)}

Based on the CRYSTALS-Dilithium \cite{crystals_dilithium} signature scheme, ML-DSA \cite{nist_mldsa} is a post-quantum digital signature algorithm designed to replace classical schemes like RSA and ECDSA. Being a signature scheme, it supports the standard \textbf{KeyGen, Sign, and Verify} operations, and is defined across three parameter sets—ML-DSA-44, ML-DSA-65, and ML-DSA-87—with \textit{ML-DSA-65} offering approximately 192-bit classical security, making it well-suited for applications such as O-RAN environments.  \textbf{HashML-DSA} is another standardized version of the algorithm.

ML-DSA is built upon two core hardness assumptions: the Module Learning With Errors (MLWE) problem and the Module Short Integer Solution (MSIS) problem, both of which are believed to be resistant to attacks even by quantum computers. Specifically, MLWE generalizes the standard Learning With Errors (LWE) problem, where the goal is to distinguish between valid LWE samples and uniformly random noise (D-LWE). ML-DSA parameter sets are shown in Table~\ref{tab:mldsa-sizes}.

\begin{table}[h!]
\centering
\footnotesize
\setlength{\tabcolsep}{4pt}
\renewcommand{\arraystretch}{1.2}
\begin{tabular}{lccc}
\rowcolor{lightblue}
\textbf{Set} & \textbf{Pub (B)} & \textbf{Sec (B)} & \textbf{Sig (B)} \\
ML-DSA-44 & 1,312 & 2,560 & 2,420 \\
ML-DSA-65 & 1,592 & 4,032 & 3,309 \\
ML-DSA-87 & 2,592 & 4,896 & 4,627 \\
\end{tabular}
\caption{ML-DSA parameter set sizes (bytes)}
\label{tab:mldsa-sizes}
\end{table}

\noindent
The type and security of Post-Quantum Schemes are summarized in Table~\ref{tab:pq_schemes}.

\begin{table*}[h!]
\centering
\small
\rowcolors{2}{lightgray}{white}
\begin{tabular}{>{\bfseries}m{3.2cm} m{3.2cm} m{3.2cm} m{3.2cm}}
\rowcolor{lightblue}
\textbf{Scheme} & \textbf{Type} & \textbf{Security Goal} & \textbf{Security Level (NIST)} \\
ML-KEM     & KEM (Lattice) & IND-CCA & Level 1–5 \\
ML-DSA (Dilithium) & Signature (Lattice) & SUF-CMA & Level 1–5 \\
SLH-DSA & Signature (Hash-based) & EUF-CMA & Level 1–5 \\
NTRU & KEM (Lattice) & IND-CCA & Level 1–5 \\
X-Wing & KEM (Structured Lattice) & IND-CCA & Level 1–5 \\
Composite-Signatures & Hybrid & SUF-CMA/EUF-CMA (composite) & Variable \\
\end{tabular}
\caption{Security Properties of Selected PQ Schemes}
\label{tab:pq_schemes}
\end{table*}

\subsubsection{Quantum Random Number Generators (QRNGs)}
Quantum Random Number Generators (QRNGs) extract entropy from inherently probabilistic quantum processes—such as photon arrival times, quantum vacuum fluctuations, or single-photon path superposition—to generate truly random bitstreams. Unlike pseudo-random number generators (PRNGs), which are algorithmically deterministic and thus potentially vulnerable to state compromise, QRNGs offer information-theoretic security.

This makes them particularly well-suited for post-quantum cryptography (PQC), where classical entropy sources may not provide sufficient resistance against quantum-enabled adversaries. High-entropy QRNG outputs are used to seed key generation, key encapsulation/encryption, and signature randomization algorithms, ensuring robust entropy pools and resistance against side-channel entropy prediction. 

\begin{figure}[h]
    \centering
    \includegraphics[width=0.45\textwidth]{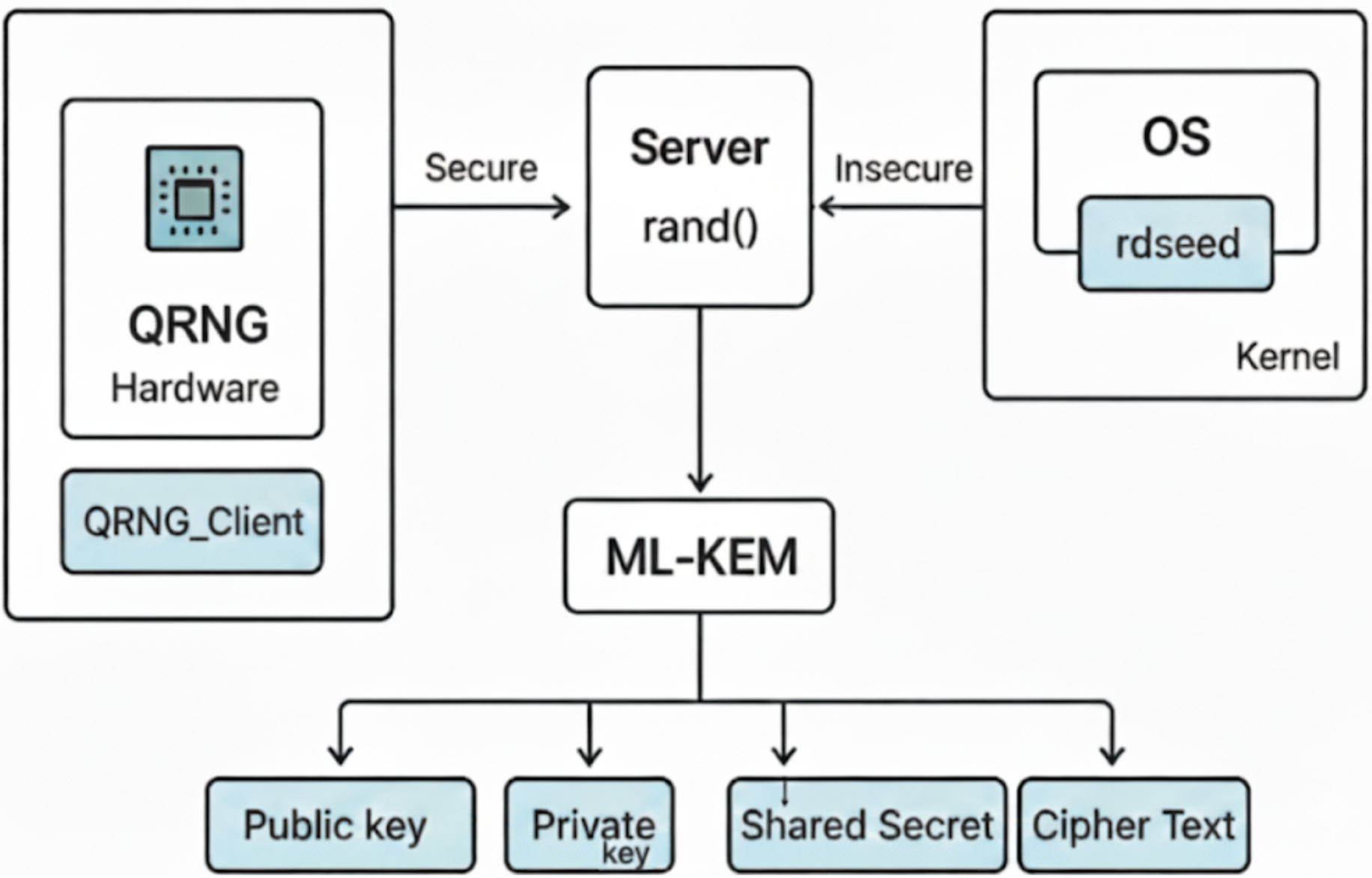}
    \caption{QRNG-based key generation feeding ML-KEM with true quantum entropy.}
    \label{fig:qrng}
\end{figure}

In order to achieve FIPS 140-3 \cite{fips_140_3} compliance, a QRNG chip must be integrated using a FIPS specified RNG method and follow the rigorous and strict criteria for entropy and random number generation outlined in the FIPS standard. The integration shall also comply with NIST SP 800-90A \cite{nist_800_90a}---requirements for random number generation. An example QRNG method combining entropy from two sources, for use in ML-KEM, is illustrated in the Figure~\ref{fig:qrng}.

\subsubsection{Quantum-Resistant Symmetric Cryptography}

AES-256 remains secure against Grover’s \cite{grover} algorithm with an effective 128-bit post-quantum strength \cite{nist_aes_report}. SHA-3\cite{fips_202} retains its collision and pre-image resistance properties, making it suitable for post-quantum cryptography (PQC) use cases such as hashing, digital signatures, and authentication.

SHA-3 is built using a sponge construction that incorporates the Keccak-f[1600] permutation as its core, along with additional steps such as domain separation and output truncation. In addition to the fixed-output hash functions, SHA-3 includes two extendable output functions (XOFs)—SHAKE128 and SHAKE256—which are now used in NIST FIPS 203 and 204 for internal hashing.

The algorithm is also well-suited for hardware acceleration, leveraging instruction sets such as AVX2 on x86 and NEON on ARM-based platforms.

\subsubsection{Post-Quantum TLS/DTLS}

\begin{figure*}[t]
    \centering
    \includegraphics[width=0.5\textwidth]{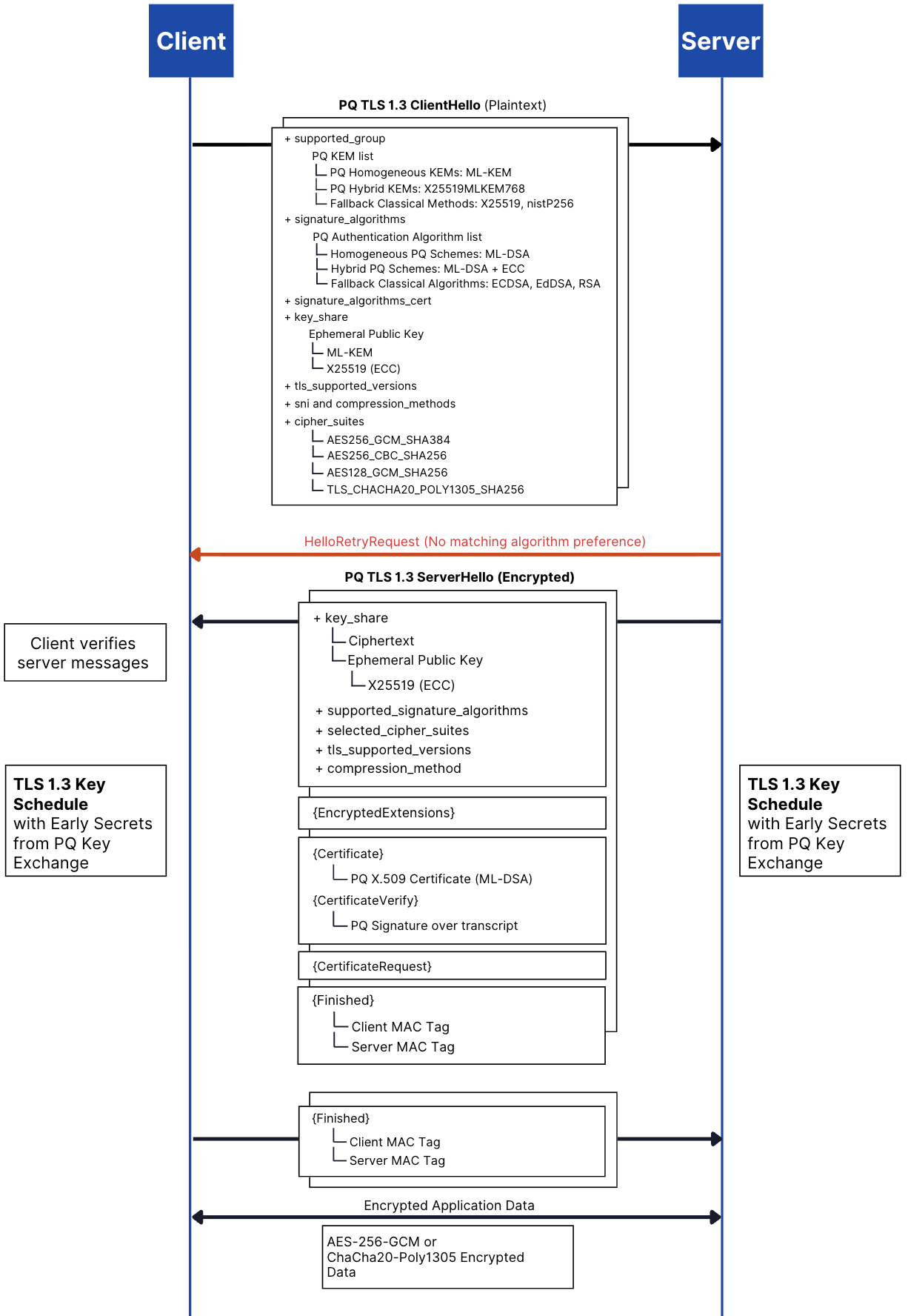}
    \caption{Post-Quantum TLS handshake flow with PQ KEM and authentication.}
    \label{fig:pq_tls_flow}
\end{figure*}

PQ-TLS and PQ-DTLS integrate post-quantum and classical algorithms through hybrid handshakes, ensuring crypto-agility and backward compatibility while providing quantum resistance. These protocols allow for a secure migration path without compromising performance.
The integration is illustrated in Figure~\ref{fig:pq_tls_flow}.

\subsubsection{Post-Quantum IPsec}

PQ-IPsec is an extension to the classical IKEv2, which integrates Post-Quantum Hybrid KEMs, such as ECDHE+ML-KEM-768 for shared key establishment, simultaneously mixing Post-Quantum Pre-Shared Keys (PPKs, RFC~8784). Thus, it provides a mitigation to the Harvest-Now-Decrypt-Later risks and enhances the security of IPsec tunnel establishment.

\subsubsection{Post-Quantum Certificate Authority (PQ-CA)}

A PQ-CA issues and validates certificates using PQ signatures, such as ML-DSA / SLH-DSA, replacing RSA/ECC-based roots of trust. PQ-CAs provide long-term authenticity for devices, applications, and xApps/rApps within hierarchical O-RAN trust domains.

\section{Migration to Quantum-Secure O-RAN}

This section presents a comprehensive strategy for migrating O-RAN networks from classical to quantum-resistant security. The migration aims to address every interface, protocol, and component in the O-RAN architecture, ensuring a consistent and practical strategy feasible for network operators.

\subsection{Migration Overview}

% Table \ref{tab:migration} provides a comprehensive mapping of O-RAN interfaces, their current classical security mechanisms, and required quantum-secure replacements. This systematic approach ensures no interface remains vulnerable to quantum attacks.

A concise summary of O-RAN interfaces, their current classical security mechanisms, and required quantum-secure replacements is provided in Table~\ref{tab:migration}. This systematic approach guarantees that no interface remains vulnerable to quantum attacks.

\renewcommand{\arraystretch}{1.2}
\small
\setlength{\tabcolsep}{4pt}

\begin{table*}[!t]
\centering
\caption{Classical O-RAN to Post-Quantum O-RAN Migration}
\label{tab:migration}
\begin{tabular}{>{\raggedright}p{2.5cm} >{\raggedright}p{4cm} >{\raggedright}p{3cm} >{\raggedright\arraybackslash}p{3.5cm}}
\toprule
\rowcolor{lightblue}
\textbf{Phase / Interface} & \textbf{Between Nodes} & \textbf{Existing Security} & \textbf{Quantum-Secure Mechanisms} \\ 
\midrule
\rowcolor{lightgray}
MidHaul - F1AP & O-CU \& O-DU; O-CU-CP \& O-DU (F1-C); O-CU-UP \& O-DU (F1-U) & IPsec or DTLS & PQ-IPsec or PQ-DTLS \\ 
\midrule
E1AP & O-CU-CP \& O-CU-UP & IPsec or DTLS & PQ-IPsec or PQ-DTLS \\ 
\midrule
\rowcolor{lightgray}
BackHaul - NGAP & O-CU \& AMF; O-CU-CP \& 5G Core & IPsec or DTLS & PQ-IPsec or PQ-DTLS \\ 
\midrule
E2 & Near-RT RIC \& O-CU; Near-RT RIC \& O-DU & IPsec or DTLS & PQ-IPsec or PQ-DTLS \\ 
\midrule
\rowcolor{lightgray}
Xn & Source gNB \& Target gNB & IPsec or DTLS & PQ-IPsec or PQ-DTLS \\ 
\midrule
BackHaul N3 & O-CU \& UPF; O-CU-UP \& UPF & IPsec & PQ-IPsec \\ 
\midrule
\rowcolor{lightgray}
O-FH (M-Plane) & O-RU \& O-DU/SMO & mTLS, SSHv2 & PQ-mTLS or PQ-SSH \\ 
\midrule
A1 & Near-RT RIC \& Non-RT RIC & mTLS & PQ-mTLS \\ 
\midrule
\rowcolor{lightgray}
O1 & SMO \& O-RAN Managed Elements & mTLS & PQ-mTLS \\ 
\midrule
O2 & SMO \& O-Cloud & mTLS & PQ-mTLS \\ 
\midrule
\rowcolor{lightgray}
Y1 & Near-RT RIC to authorized consumers & mTLS & PQ-mTLS \\ 
\midrule
R1 & rApps \& Non-RT RIC framework & mTLS & PQ-mTLS \\ 
\bottomrule
\end{tabular}
\end{table*}

Prior research, such as \cite{OteroGarcia2025}, has addressed post-quantum security enhancements for specific protocols like IPsec and MACsec in RAN contexts. This work, however, presents a complete and practical migration strategy, covering all O-RAN interfaces and components to enable a seamless transition from classical to quantum-resistant security across the entire architecture.

\subsection{Foundational Entropy: QRNG Integration}

% QRNGs form the foundational security layer for all quantum-resistant cryptographic operations. Integrating them throughout the O-RAN architecture ensures that cryptographic keys are generated with true, verifiable randomness.

Strong cryptographic security is not just limited to the underlying hard problems but a variety of other supporting factors, with a quality, unpredictable, and safe randomness source being an equally important criterion. QRNGs, being an excellent source of true randomness, fulfill this requirement, and with their integration into cryptographic algorithms, we ensure the operations of keygen, encapsulation, and (optionally)signing are truly random.

\subsubsection{QRNG Deployment Architecture}

The QRNG infrastructure consists of hardware QRNGs strategically placed within the network architecture. A primary QRNG is provisioned within the SMO, which provides high-entropy randomness for the PQ-CA's root key generation and other certificate issuance/signing operations. The QRNGs utilized have undergone rigorous NIST's Entropy Validation Programs.

Network functions incorporate QRNGs in their local entropy pools, accessing them via FIPS 140-2/3 validated cryptographic modules, for use in TLS connections, long-lived IPsec sessions, and other cryptographic operations. The QRNG chips are typically connected via PCIe, and the random bitstream is processed inside secure memory enclaves---such as those provided by Intel SGX, or AMD SEV.

\subsubsection{QRNG Applications in O-RAN}

\textbf{Securing Open Interfaces:} O-RAN’s disaggregated architecture exposes multiple open interfaces—Open Fronthaul, F1, E2, A1, O1, and Xn—that need robust protection. QRNGs generate keys for post-quantum protocols such as PQ-IPsec, PQ-mTLS, and PQ-DTLS, ensuring confidentiality, integrity, and authentication for all data passing through these interfaces.

\textbf{RIC Security:} The RIC is central to controlling and optimizing the RAN, making its security critical. QRNGs provide cryptographic keys for securing the RIC platform, its communications, and AI/ML-driven applications (xApps and rApps). This includes securing communications over the A1 interface (Non-RT RIC to Near-RT RIC) and the E2 interface (Near-RT RIC to RAN nodes).

\textbf{Certificate Generation and PKI:} O-RAN's Zero Trust Architecture \cite{oran_zta} relies on certificate-based mutual authentication. QRNGs provide unpredictable randomness for generating private keys for post-quantum X.509 certificates, establishing a root of trust for all network functions (O-RU, O-DU, O-CU, RICs).

\textbf{Secure Application Lifecycle:} xApp and rApp integrity is critical. QRNGs generate keys for digital signatures that sign and verify software packages during application onboarding and lifecycle management, preventing deployment of malicious or compromised code.

\textbf{Data at Rest Protection:} RICs and management components store sensitive data including network policies, performance metrics, and configuration data. QRNGs generate symmetric encryption keys for encrypting this data at rest, protecting it from unauthorized access even if components are compromised. The keys are stored in a Hardware Security Module (HSM) provisioned at the RIC.

\subsection{Control Plane Security: AES-128 to AES-256 Upgrade}

The Radio Resource Control (RRC) layer handles the crucial control-plane signaling between the User Equipment (UE) and the gNB. Since it carries sensitive control messages (e.g., UE Identifiers and capabilities), it needs to be well protected against eavesdropping and tampering. To do this, RRC currently uses 128-bit algorithms like NIA and NEA (for example, AES-128 and SNOW-3G). If these protections are broken, attackers could interfere with control data, disrupt handovers, or even launch Denial-of-Service (DoS) attacks. However, with the rise of quantum threats, the 128-bit security level is no longer enough. Moving to stronger algorithms like AES-256 is necessary, but this shift will require coordinated updates across the NAS (Non-Access Stratum), the UE itself, and the 5G Core network.

The mathematical justification for the mentioned update lies in defense against Grover's algorithm, which provides square-root speedup over classical exhaustive key-search attacks. For a symmetric cipher with a k-bit key, classical key-search requires approximately $2^k$ operations, while Grover's algorithm reduces this to approximately $2^{k/2}$ quantum queries.

For AES-128, quantum attack effective security drops from 128 bits to 64 bits ($2^{128}$ classical operations becomes $2^{64}$ quantum queries). A 64-bit security level is vulnerable to sufficiently powerful quantum computers. For AES-256, effective security under Grover's algorithm drops from 256 bits to 128 bits ($2^{256}$ classical operations becomes $2^{128}$ quantum queries) \cite{nist_aes_report}. A 128-bit security level remains computationally infeasible even for quantum computers, providing durable, long-term security.

% This transition ensures that critical control-plane signaling—such as handover commands, connection setups, and resource allocations—remains confidential and secure against advanced quantum threats.
The move to AES-256 applies not only to RRC traffic but to all symmetric encryption operations across the O-RAN architecture, with extensions to 5G core and UE in the future.

\subsection{User and Control Plane Interfaces: PQ-IPsec and PQ-DTLS}

\begin{figure*}[t]
    \centering
    \includegraphics[width=0.9\textwidth]{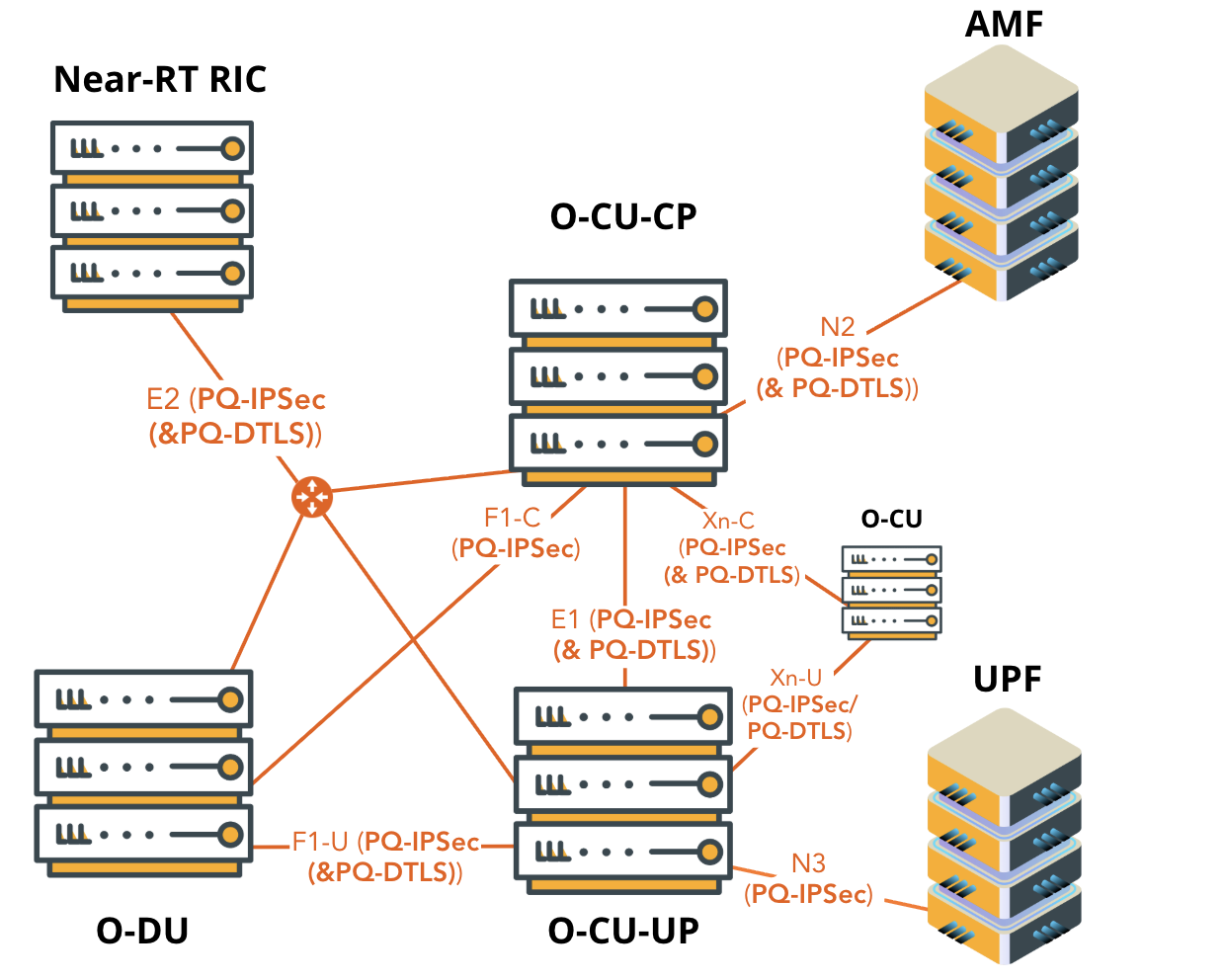}
    \caption{O-RAN Architecture with Security Protocols showing PQ-IPsec(\& PQ-DTLS) on various interfaces (F1-C, F1-U, E1, Xn-C, Xn-U, N2, N3, E2) connecting Near-RT RIC, O-CU-CP, O-CU-UP, O-DU, AMF, and UPF.}
    \label{fig:oran_security}
\end{figure*}

The security (confidentiality and authentication) of interfaces like F1AP, E1AP, Xn, N2, N3, and E2 is typically provided by IPsec or DTLS, both of which are still susceptible to quantum attacks. Upgrading these protocols to PQ-IPsec and PQ-DTLS, which incorporate post-quantum cryptographic algorithms and QRNG-generated entropy, is necessary to achieve quantum-resistant security.
\noindent\\
Figure~\ref{fig:oran_security} illustrates the deployment of PQ-IPsec and PQ-DTLS across O-RAN interfaces.

\subsubsection{PQ-IPsec Implementation Strategy}

The PQ-IPsec implementation modifies the Internet Key Exchange version 2 (IKEv2) protocol to support PQC for key establishment through a hybrid approach combining classical key exchange (ECDH) with PQ Key Encapsulation Mechanism (ML-KEM).

\textbf{Hybrid Key Exchange:} The IKE\_SA\_INIT exchange establishes a secure channel using classical algorithms (e.g., ECDH with Curve 25519). Subsequently, ML-KEM-768 key exchange occurs within an encrypted IKE\_INTERMEDIATE message. This two-step process accommodates large PQC key sizes that cannot be fragmented in initial handshakes. The encrypted nature of IKE\_INTERMEDIATE provides traffic flow confidentiality.

\textbf{Post-Quantum Pre-shared Keys:} RFC 8784 defines Post-Quantum Pre-shared Keys (PPKs) providing an additional quantum resistance layer. High-entropy, quantum-generated pre-shared keys (256 bits classical security) are provisioned out-of-band to communicating peers. These PPKs are mixed into key derivation functions (KDFs) which ensures that even if adversaries record entire IKEv2 exchanges and later break asymmetric cryptography with quantum computers, they cannot derive final session keys without PPKs, rendering harvested data useless.

\textbf{QRNG for PPK Generation:} PPK security depends entirely on unpredictability. QRNGs generate PPKs ensuring highest possible entropy, making them truly random secrets immune to prediction.

Table~\ref{tab:pqtls-ipsec} provides the specifications for Post-Quantum TLS and IPsec.

\subsubsection{PQ-DTLS Implementation Strategy}

\begin{figure*}[htbp]
    \centering
    \includegraphics[width=0.8\textwidth]{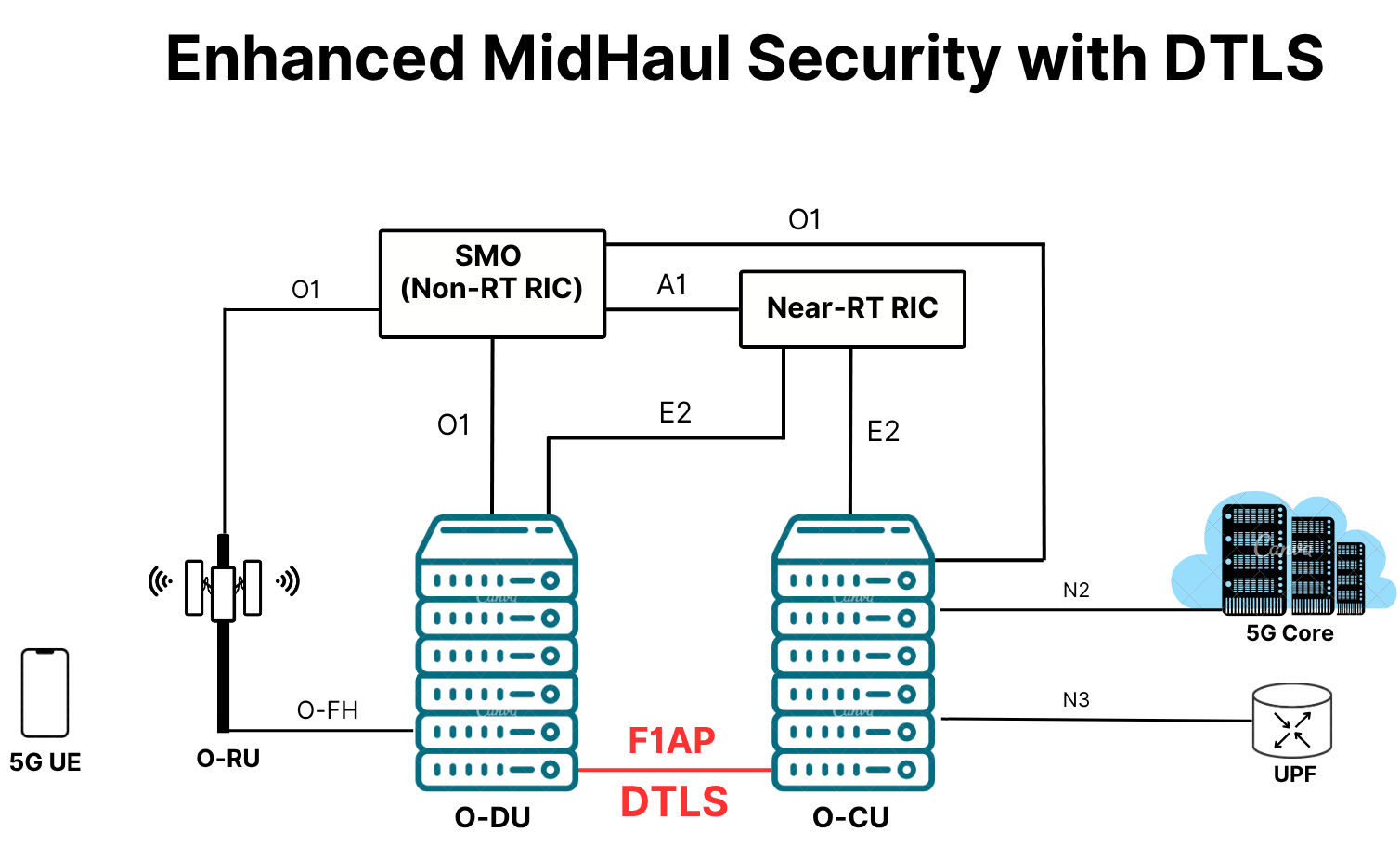}
    \caption{PQ-DTLS 1.3 Protocol at F1AP Interface}
    \label{fig:pq-dtls}
\end{figure*}

For interfaces using DTLS over SCTP \cite{rfc_8261} (F1AP, N2, E1AP), migration to quantum-safe posture involves upgrading to PQ-DTLS 1.3 with multiple quantum-safe enhancements. Figure~\ref{fig:pq-dtls} shows the PQ-DTLS 1.3 protocol over the F1AP interface (DU-CU).

\textbf{Upgraded Cipher Suites:} Handshakes are fortified by replacing classical key exchange with ML-KEM. For symmetric encryption, cipher suites upgrade to AES-256, providing 128-bit security against Grover's algorithm attacks.

\textbf{PQ Signature Schemes:} Authentication becomes quantum-safe by replacing traditional signatures (RSA, ECDSA) with PQC signature schemes. This involves issuing and validating PQ X.509 certificates signed with ML-DSA.

\textbf{QRNG for Keys and Nonces:} PQC scheme security critically depends on true randomness. QRNGs provide high-entropy seeds for generating ML-KEM secret keys and unpredictable nonces for ML-DSA signature operations, preventing attacks exploiting predictable or repeated random values.

\begin{figure*}[h!]
    \centering
    \includegraphics[width=0.6\textwidth]{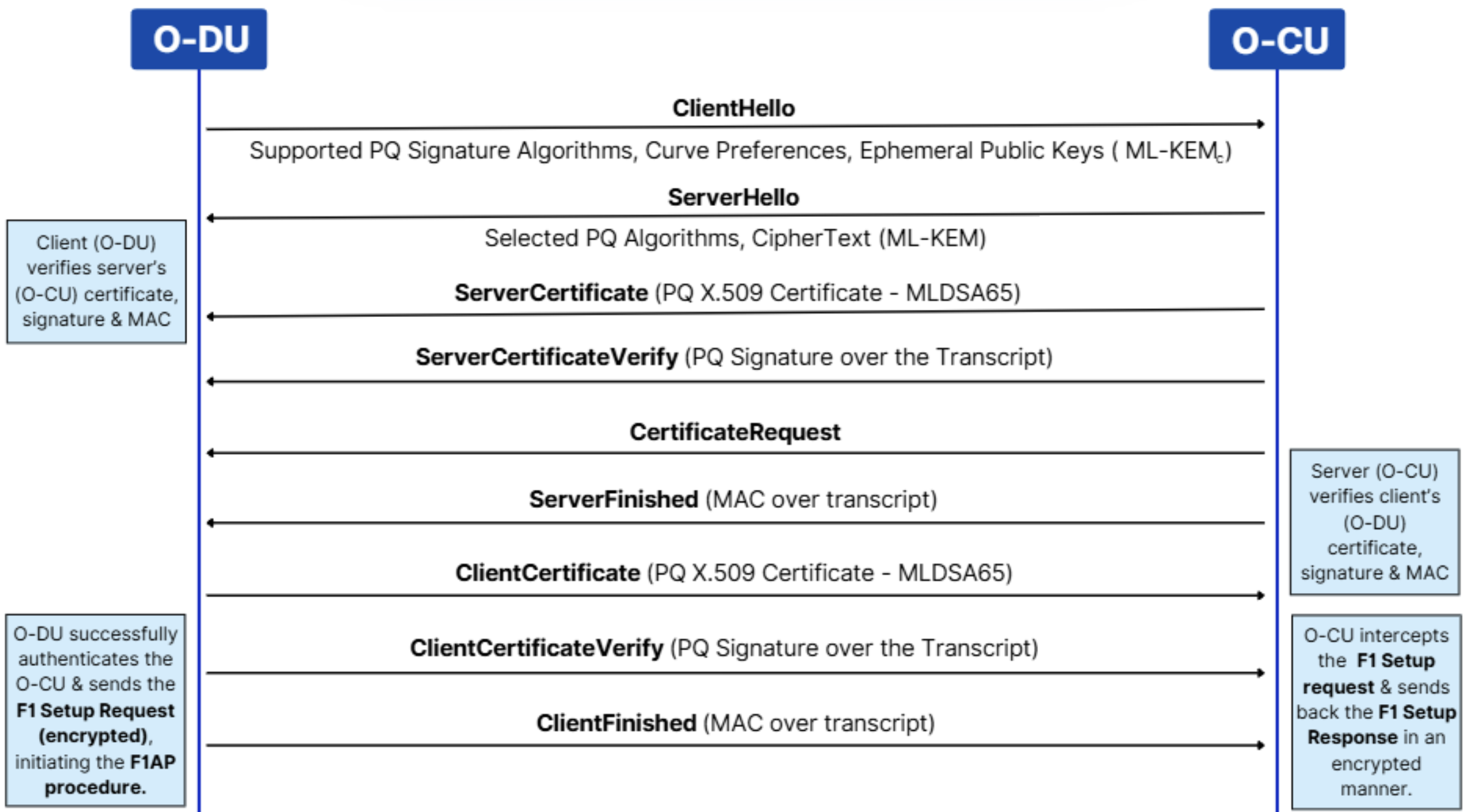}
    \caption{Post-Quantum DTLS 1.3 Handshake Flow between O-DU and O-CU showing ML-KEM-based key exchange and ML-DSA-65 signed X.509 certificates for mutual authentication, ensuring a quantum-resilient F1 interface setup.}
    \label{fig:pq_dtls}
\end{figure*}

\textbf{Figure~\ref{fig:pq_dtls}} illustrates the PQ-DTLS 1.3 handshake sequence between O-DU and O-CU. It demonstrates the exchange of PQ X.509 certificates (ML-DSA-65), mutual verification of signatures, and encrypted session establishment using ML-KEM, achieving end-to-end post-quantum secure communication over the F1 interface.

\subsection{Management Interfaces: PQ-mTLS Deployment}

\begin{figure*}[t]
    \centering
    \includegraphics[width=0.5\textwidth]{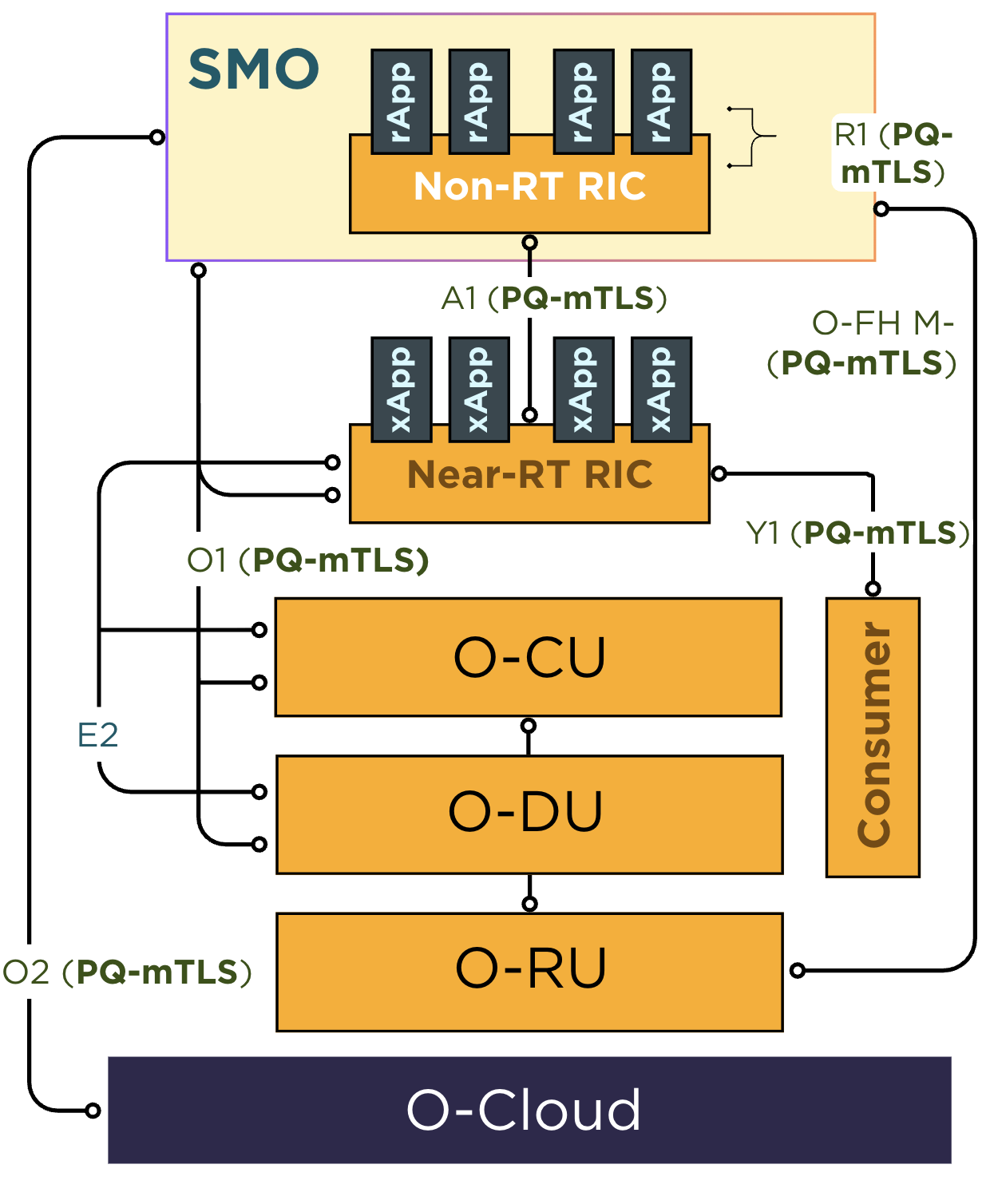}
    \caption{O-RAN SMO Architecture with Security Protocols showing PQ-mTLS on interfaces (R1, A1, O1, O2, O-FH M-Plane, Y1) and PQ-OAuth2.0 for rApps, xApps, and Y1 Consumers.}
    \label{fig:smo_security}
\end{figure*}

Management and control interfaces (A1, O1, O2, Y1, R1) facilitate communication between essential components including the SMO, RICs, and O-Cloud. Traditionally secured using mTLS, these interfaces must be upgraded to Post-Quantum mTLS (PQ-mTLS) which integrates quantum-resistant algorithms (e.g., ML-KEM and ML-DSA) into TLS 1.3 handshakes and certificate infrastructure.
\noindent
Figure~\ref{fig:smo_security} illustrates PQ-mTLS deployment across management interfaces.

\subsubsection{PQ-mTLS Implementation}

Securing API-driven management interfaces involves upgrading to PQ-mTLS, applying PQC algorithms to TLS 1.3 handshakes with focus on mutual authentication and backward compatibility.

\textbf{Hybrid Key Exchange:} Implementation adopts hybrid cipher suites. During TLS 1.3 handshakes, classical key exchange (ECDHE) and PQC KEM (ML-KEM) run in parallel. Final shared secrets derive from both computations, ensuring connection security as long as at least one algorithm remains unbroken.

\textbf{Mutual Authentication with PQ Certificates:} Quantum-resistant mutual authentication requires both entities to exchange and validate PQ Certificates signed with ML-DSA. This ensures, for example, that Non-RT RIC and Near-RT RIC can cryptographically verify each other's identity before establishing connections.

\textbf{QRNG-Hardened Key Generation:} QRNGs provide truly random seeds for all sensitive cryptographic material, including ephemeral keys for each ML-KEM exchange and long-term private keys for PQ certificates.

\renewcommand{\arraystretch}{1.5}
\begin{table}[h!]
\centering
\small
\rowcolors{2}{lightgray}{white}
\begin{tabular}{>{\bfseries}m{2.6cm} m{5cm}}\rowcolor{lightblue}
\textbf{Protocol} & \textbf{PQ TLS /IPsec Specification} \\
PQ TLS 1.3 & Hybrid KEM (X25519 + ML-KEM) for key exchange \\
PQ IPsec  & PQ SA negotiation using composite KEM proposals \\
Cipher Suites & AES-GCM / ChaCha20-Poly1305 + SHA-384 \\
Cert Format & X.509 with PQ SubjectPublicKeyInfo extensions \\
Handshake & Post-Quantum + Classical hybrid authentication \\
\end{tabular}
\caption{Post-Quantum TLS and IPsec Specification Overview}
\label{tab:pqtls-ipsec}
\end{table}
\subsection{Centralized Trust: PQ-CA and PQ-OAuth in SMO}

The SMO's central management and automation role makes it the logical anchor for establishing network-wide root of trust. Upgrading the SMO to host a Post-Quantum Certificate Authority (PQ-CA) establishes the cornerstone of Post-Quantum PKI, issuing and managing cryptographic identities for all components and applications—a foundational requirement for O-RAN's Zero Trust Architecture \cite{oran_zta}. The CA specifications are summarized in the Table~\ref{tab:ca}.

\begin{table*}[h!]
\centering
\small
\rowcolors{2}{lightgray}{white}
\begin{tabular}{>{\bfseries}m{3.0cm} m{5cm}}
\rowcolor{lightblue}
\textbf{Component} & \textbf{PQ Certificate Authority (CA) Specification} \\
Root CA & Hybrid keypair (Ed448/p384 + ML-DSA-65) \\
Certificate Profile & Composite or hybrid public key + PQ extensions \\
OCSP / CRL  & Signed with PQ-compatible algorithm \\
Key Storage & Hardware Security Modules \\
Key Usage & CA Signature, CRL Sign \\
\end{tabular}
\caption{Post-Quantum Certificate Authority Specification}
\label{tab:ca}
\end{table*}

\subsubsection{Post-Quantum Certificate Lifecycle Management}

The Post-Quantum Certificate Authority (PQ-CA), deployed within the SMO, manages the full lifecycle of quantum-resistant digital certificates for O-RAN components, covering generation, issuance, distribution, renewal, and revocation.

\textbf{Certificate Generation and Hierarchy:} The PQ-CA issues quantum-resistant X.509 certificates for all critical O-RAN entities including Non-RT RIC, Near-RT RIC, rApps, and xApps. The PKI structure follows a standard hierarchy with a PQ Root Certificate at the top, one or more PQ Intermediate CAs, and PQ Leaf Certificates for individual devices and applications. Certificates are signed using quantum-secure signature schemes such as ML-DSA, ensuring signatures remain secure against quantum adversaries.

\textbf{Key Material Security via QRNG:} To guarantee cryptographic strength, the PQ-CA generates private keys using a Quantum Random Number Generator (QRNG). This source of true quantum entropy strengthens the unpredictability of keys, essential for resisting brute-force attacks both now and post-quantum era.

\textbf{Certificate Renewal and Revocation:}

\textit{Renewal}: Certificates are issued with short validity periods (7 to 90 days) to limit exposure in case of key compromise. The PQ-CA supports automated renewal workflows triggered before expiration, minimizing service disruption.

\textit{Revocation:} Certificate revocation relies on digital records like Certificate Revocation Lists (CRLs) and Online Certificate Status Protocol (OCSP) responders, which have been extended to support post-quantum certificates. These mechanisms allow operators to verify certificate status in real time and revoke certificates promptly, which is critical for maintaining security in fast-changing 5G environments.

\textbf{Automated Management Protocols:} To handle certificate lifecycles at scale, the ACME protocol \cite{rfc_8555} has been adapted for post-quantum cryptography. This allows for automatic issuance, renewal, and revocation of PQ X.509 certificates without manual steps. ACME protocol sessions are protected via PQ-mTLS, with regular re-keying to mitigate nonce reuse risks in GCM modes.

\textbf{Operational Considerations:} The post-quantum Certificate Authority (PQ-CA) integrates directly with O-RAN orchestration and monitoring tools, i.e., the SMO. It offers endpoints for provisioning and monitoring certificates, providing easily accessible certificate lifecycle monitoring functionalities to various participating entities.

\subsubsection{Quantum-Secure Authorization for RIC Applications}

In multi-vendor environments with third-party applications, correct (and secured) authentication and authorization are critical. The SMO, enhanced with PQ-CA, and a PQ OAuth 2.0 Server (AS) functions as a quantum-secure authorization server managing access control for rApps and xApps.

\begin{figure}[!h]
    \centering
    \includegraphics[width=0.8\linewidth]{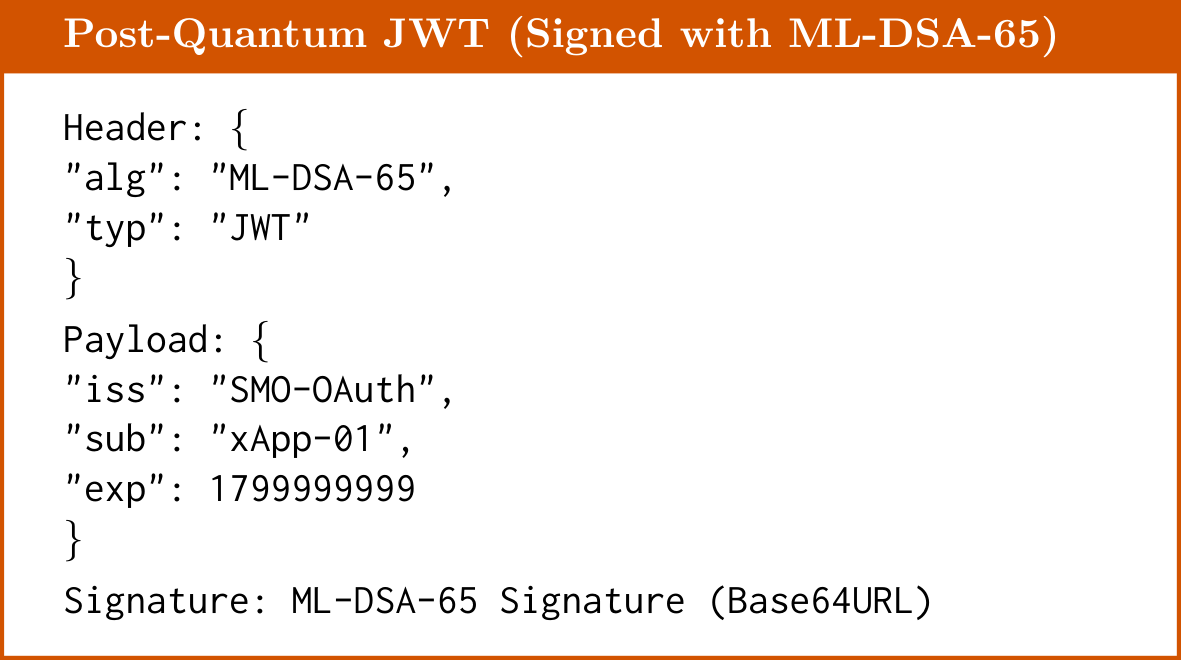}
    \caption{Post-Quantum JWT Example}
    \label{fig:pq-jwt}
\end{figure}

\textbf{PQ-Token Based Authorization:} When rApps or xApps need access to protected resources or APIs (e.g., over R1 or A1 interfaces), they request access tokens from the SMO. The SMO issues Post-Quantum Tokens (PQ-Tokens), typically JSON Web Tokens (JWT) where signatures are generated using ML-DSA, making tokens verifiable and tamper-proof against quantum adversaries. A sample PQ-Token is shown in Figure~\ref{fig:pq-jwt}.

\textbf{Secure Communication Channels:} Communication channels used by SMO components, RICs, and applications are secured using PQ-mTLS. Mutual authentication for connections uses PQ certificates issued by PQ-CA. This ensures when PQ-Tokens are transmitted, they travel over channels that are confidentially protected and authenticated with quantum-resistant identities, creating comprehensive, end-to-end secure frameworks.

\section{Implementation of PQ-Security in Open RAN}

Our Post-Quantum RAN security solution was implemented by extending the cryptographic and transport layers of the Open RAN software stack, using both standardized and experimental libraries that support quantum-resistant algorithms.

\subsection{O-RAN Software Stack}

To explore the integration of Post-Quantum security protocols, we adopted an O-RAN-compliant software stack. We chose \textbf{OpenAirInterface5G (OAI)}~\cite{oai_ran} as the base for our disaggregated gNB (CU/DU) stack, given its commendable performance and close alignment with 3GPP and O-RAN specifications. We plan to extend this Post-Quantum framework to the O-RAN Distributed Unit (DU) and evaluate other open-source RAN implementations as well.

\subsection{Cryptographic Libraries}

Most current cryptographic libraries lack full support for PQC-enhanced versions of security protocols, primarily due to the rapidly evolving nature of the field. However, through custom patches and by integrating multiple libraries, we were able to achieve functional Post-Quantum support—albeit not at full production maturity. The following libraries were used to support PQC primitives:

\begin{itemize}
    \item \textbf{liboqs}~\cite{liboqs}: A C library providing quantum-safe Key Encapsulation Mechanisms (KEMs) and digital signature algorithms. It was used to enable PQ support in TLS handshakes and certificate verification across TLS/DTLS protocols.
    
    \item \textbf{OpenSSL}~\cite{openssl}: One of the most widely used cryptographic libraries, supporting TLS/DTLS and certificate management. Recent versions offer support for ML-KEM-based hybrid key exchanges in TLS 1.3, which we utilized in our mTLS and DTLS implementations.
    
    \item \textbf{strongSwan}~\cite{strongswan}: A robust and widely deployed IPsec/IKEv2 implementation, supporting advanced features like EAP-AKA, MOBIKE, and Traffic Confidentiality. Since version 6.0.0, strongSwan supports the Post-Quantum IPsec extensions defined in RFC 8784 and RFC 9370.
    
    \item \textbf{wolfSSL}~\cite{wolfssl}: A lightweight TLS/DTLS implementation with native support for PQ algorithms and TLS 1.3. We used wolfSSL to validate PQ-DTLS integration at the RAN layer and to gain additional debugging visibility.

    \item \textbf{cuPQC} ~\cite{cupqc}: A CUDA-based library for accelerating PQ primitives on NVIDIA GPUs, offering significant performance increases in comparison to CPU-based or software-level implementations. This library is especially useful in cases where we have to serve multiple client requests using IPsec/TLS concurrently (e.g., at SMO). It provides configuration options to tune thread parameters in CUDA-kernels. \textbf{cuHash} is another library for offloading SHA-2 / SHA-3 and SHAKE-128 / 256 hash functions on the GPU, which are highly useful in speeding up KDFs, MACs, Merkle Tree based applications, etc.
    
\end{itemize}

\begin{figure}[h]
    \raggedright
    \includegraphics[width=0.4\textwidth]{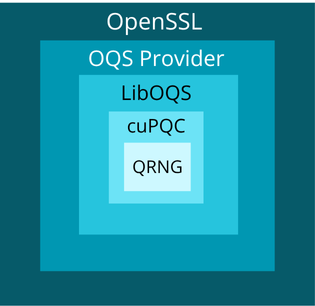}
    \caption{Cryptographic Libraries stack utilized in Q-RAN}
    \label{fig:libs-pq}
\end{figure}

Figure~\ref{fig:libs-pq} shows the layered approach to achieve PQ Security using the specified libraries.

\textbf{}

\subsection{Integration into Open RAN Stack}

\subsubsection{gNodeB: PQ-DTLS and PQ-IPsec}

Since OAI does not natively support DTLS on RAN interfaces, DTLS 1.3 was integrated manually, with built-in certificate authority (CA) support and modular configuration for certificate and protocol parameter selection across vulnerable interfaces. In this setup, DTLS 1.3 was layered on top of SCTP, similar to how TLS operates over TCP. The DTLS stack was built using OpenSSL's SSL objects—designed to work over datagram sockets—and further extended using liboqs to add support for post-quantum cryptographic primitives. Post-quantum operations were offloaded to NVIDIA GPUs using \textbf{cuPQC} — to accelerate PQC primitives — significantly improving throughput, especially for short-lived DTLS sessions that require frequent rekeying. It further helped in sustaining the desired throughput expected from 5G RAN.

The IETF Draft~\cite{draft_dtls_sctp_2} extends the DTLS 1.3 protocol for use over SCTP datagrams, adding support for SCTP-AUTH chunks, multihoming, multiple streams, and preserving SCTP’s built-in reliability mechanisms. However, most open-source TLS/DTLS libraries do not natively support these features. As a result, a key challenge was ensuring interoperability between DTLS 1.3 implementations and OAI’s native SCTP interfaces—without compromising any existing SCTP functionality.

The encryption overhead added by DTLS also had to be carefully optimized to closely match the throughput of unencrypted links. Additionally, IKEv2 and IPsec ESP were configured in tunnel mode via strongswan to guarantee confidentiality and authentication for payloads, as well as the endpoint metadata. Rekeying intervals were appropriately set up to maintain forward secrecy and ensure connection freshness. To sustain the high-throughput requirements on the 5G data plane, IPsec ESP processing was offloaded to NVIDIA BlueField-3 (BF3) DPUs using the DOCA libraries~\cite{doca}, enabling performance levels suitable for production-grade deployments.

\subsubsection{SMO: PQ-mTLS and PQ-OAuth 2.0}

Post-Quantum mTLS 1.3 was integrated into the O-RAN's SMO and RIC components using a dedicated service mesh enhanced with support for PQ algorithms. This allowed us to deploy primitives such as ML-KEM-768 and ML-DSA-65. A Post-Quantum OAuth 2.0 server was deployed at the SMO, issuing and verifying PQ-signed tokens over interfaces like A1 and R1. Other control-plane elements were connected to each other using PQ-IPsec via strongSwan, ensuring consistent security across the RAN management domain.

\subsubsection{TLS Handshake Modes and Algorithms}

Our implementation supports both hybrid and pure Post-Quantum TLS 1.3 handshakes:
\begin{itemize}
    \item \textbf{Hybrid KEMs}: ECDHE combined with Kyber or ML-KEM (FIPS 203).
    \item \textbf{Signature Algorithms}: Ed448-ML-DSA-65, SLH\_DSA\_SHA2\_256F, Falcon\_512/1024,  other composite or pure PQ signature schemes.
    \item \textbf{Fallback and Interoperability}: The system falls back to a classical configuration using ECDHE and EdDSA/ECDSA when it detects that a peer does not support PQ-TLS 1.3 (typically when clients do not send PQ group preferences in ClientHello).
\end{itemize}

\subsection{Deployment and Validation}

The Post-Quantum secured O-CU/O-DU, SMO, and other components were deployed in a complete integrated stack (including 5G Core and Radio Units), validated against multiple end-to-end UE tests with tests for internet speed, and stability. The Post-Quantum nature of the services was verified as follows: 

\begin{itemize}
    \item \textbf{Interoperability}: Verified against Kubernetes components and classical TLS clients by integrating fallback certificate paths.
    \item \textbf{Performance}: Handshake times and encryption performance were logged and compared to classical baselines.
    \item \textbf{Security}: Validated using wolfSSL-based TLS sniffers, OpenSSL test clients for hybrid (D)TLS 1.3 draft extensions.
\end{itemize}

\section{Challenges in Migrating to Post-Quantum O-RAN Security} %% NEW

Despite standardized PQC algorithms and hybrid specifications, deploying them across the O-RAN ecosystem introduces multiple technical, operational, and interoperability challenges. This section details critical issues encountered during transition to Post Quantum versions of IPsec, TLS/DTLS, OAuth 2.0, JWKS. 

\subsection{Challenges in PQ-IPsec Implementation}

\textbf{1) Use of Post-Quantum Certificates:} Modern X.509 certificate infrastructures and protocols (such as IPsec) typically handle certificates in the size range of 0.5--1.5 KB, with complete certificate chains contributing a few additional kilobytes to the authentication data exchanged between peers. 

However, with the advent of quantum computers, the need for post-quantum cryptographic (PQC) primitives has become critical. These newer primitives, such as ML-DSA, often do not integrate seamlessly into contemporary protocols due to their significantly larger sizes (10--20 KB). This increase frequently leads to practical challenges, including increased bandwidth usage, increased connection setup delays, and a higher risk of packet loss.

Practical deployments of PQC in IKEv2 must address these issues using techniques such as certificate compression through Hash and URL method specified in~\cite{rfc_7296}, and (or) IKE message fragmentation~\cite{rfc_7383}. The contrast between key and signature sizes of classical and Post-Quantum Schemes is shown in Table~\ref{tab:crypto-sizes2}.
\begin{table*}[h!]
\centering
\small
\renewcommand{\arraystretch}{1.2}
\hspace{-1cm}
\begin{tabular}{|l|>{\columncolor{lightgray}}c|}
\hline
\textbf{Scheme} & \textbf{Key + Signature Size (KB)} \\
\hline
ECDSA                   & 0.1 \\
RSA                     & 0.5 \\
Lattice-based            & 11 \\
Stateful HBS            & 15 \\
Stateless HBS           & 42 \\
ZK Proofs (ex: Picnic L1FS) & 66\\
Multivariate            & 100 \\
Supersingular Isogenies & 122 \\
Code-based              & 190 \\
\hline
\end{tabular}
\caption{Broad estimates for key and signature sizes for selected cryptographic schemes.}
\label{tab:crypto-sizes2}
\end{table*}

5G RAN systems typically have sufficient computing capabilities and networking facilities available to them, thus reducing the impact of PQC on them. However, the increased message sizes can introduce delays in time sensitive 5G Interfaces, such as F1AP, NGAP, etc.

\textbf{2) Fragmentation and Intermediate Exchange:} Post-Quantum Key Exchanges (e.g., ML-KEM-768) introduce significantly larger key and ciphertext sizes ($\sim~$2–3 KB), in contrast to the relatively small sizes of classical algorithms like ECDH. This increase in message size is a major contributor to fragmentation in IKEv2 messages, which are carried over UDP. When message sizes exceed the Path Maximum Transmission Unit (MTU) — often limited by middleboxes such as NATs and firewalls — it can result in packet loss and delays in completing the connection.

Given the inherent unreliability of UDP, such packet loss can lead to failures in the key agreement process, especially over longer network paths. \cite{rfc_7383} addresses this by introducing a mechanism to fragment large encrypted IKE messages into multiple IKE\_Fragment payloads. However, it does not support fragmentation during the initial key exchange, where the post-quantum keys themselves are exchanged. This gap is addressed by \cite{rfc_ike_intermediate} (Intermediate Exchange), which specifies a mechanism for handling fragmentation during the initial IKE\_SA\_INIT phase by introducing Intermediate Key Exchanges which are strategically used to place Post-Quantum Keys, creating a hybrid solution.

IPsec deployments over 5G interfaces require frequent re-keying to maintain forward secrecy, keep shared secrets fresh, and mitigate risks such as nonce reuse and birthday attacks associated with GCM modes~\cite{aes_gcm_issue} 
\cite{draft_aead}. The use of post-quantum key encapsulation mechanisms (KEMs) may introduce slight delays in the re-keying process, potentially impacting time-sensitive operations in existing networks.

\textbf{3) PPK Synchronization:} RFC 8784 Post-Quantum Pre-Shared Keys (PPKs) must be securely provisioned and rotated. Distribution through SMO or SDN controllers requires new key management APIs and synchronization protocols.

\textbf{4) Hardware Acceleration:}  
High-performance data processing units (DPUs), FPGAs, and cryptographic accelerators are widely employed in 5G networks to offload and accelerate IPsec and other security protocols, ensuring state-of-the-art throughput. However, most existing hardware accelerators currently lack native support for post-quantum cryptographic (PQC) algorithms. To sustain the required throughput on high-bandwidth and throughput interfaces such as N3 and E2, integration of PQC-capable hardware accelerators or FPGA implementations is essential.

\subsection{Challenges in PQ-TLS/DTLS Integration}

% \textbf{1) Handshake Latency:} PQ-DTLS 1.3 introduces 5–10 ms additional latency per handshake due to larger key encapsulation operations. This impacts control-plane signaling on F1AP and E1AP interfaces between DU–CU.

% \textbf{2) MTU and Fragmentation:} DTLS over SCTP imposes strict packet size limits; large ML-KEM ciphertexts increase fragmentation risk, potentially causing retransmission overhead.

% \textbf{3) Certificate Chain Size:} PQ certificates (3–5 KB) strain memory and bandwidth in resource-limited DUs and RUs. Compression mechanisms and short-lived certificates mitigate this but require enhanced SMO automation.

% \textbf{4) Library Support Maturity:} While OpenSSL+OQS supports PQC handshakes, stability under SCTP transport and constrained hardware remains limited. Regression testing across multi-vendor O-RAN stacks is essential.

% \textbf{5) Forward Secrecy Preservation:} Ensuring hybrid DTLS sessions maintain forward secrecy under PQC and classical compromise requires careful KDF orchestration across both algorithms.

\textbf{1) Certificate Chain Sizes:}  
% Post-Quantum X.509 certificates introduce several bottlenecks in TLS/DTLS-secured connections due to their significantly increased size -- similar to the challenges already discussed in the context of IPsec. As evaluated in \cite{sosnowski2023performance, sikeridis2020post}, post-quantum TLS 1.3 introduces larger handshake overheads compared to classical TLS.

% In TLS (e.g., TLS 1.3~\cite{rfc_8446}), large certificate payloads are transmitted via TCP, relying on segmentation and TLS record layer fragmentation \cite{eprint_tls_pq}. However, the effectiveness of this process is bounded by the initial TCP window size and the underlying congestion control algorithm. 
Post-Quantum X.509 certificates cause noticeable bottlenecks in TLS/DTLS-secured connections due to their substantially larger size---a challenge also observed in IPsec contexts. As analyzed in \cite{sosnowski2023performance, sikeridis2020post}, post-quantum variants of TLS 1.3 introduce higher handshake overheads compared to classical cipher suites.

In TLS (e.g., TLS 1.3~\cite{rfc_8446}), these larger certificate payloads are transmitted over TCP, using segmentation and TLS record-layer fragmentation \cite{eprint_tls_pq}. However, the efficiency of this process is influenced by the initial TCP congestion window and the behavior of the selected congestion control algorithm.
If the TCP window is too small—particularly during the handshake phase—transmission of large certificate chains may be throttled, resulting in increased latency and multiple round trips if the available \texttt{init\_cwnd} or Maximum Segment Size is insufficient to transmit the full message. This is particularly impactful in high-latency paths or edge-cloud architectures where TCP slow start dominates short-lived control-plane sessions.

DTLS (e.g. DTLS 1.3~\cite{rfc_9147}), operating over UDP/SCTP, handles large handshakes by fragmenting handshake messages at the protocol level. Each fragment is acknowledged independently, and losses trigger retransmission of specific fragments, adding protocol overhead and increasing handshake latency—particularly over unreliable or constrained networks.

To mitigate certificate size overhead, TLS certificate compression has been standardized in~\cite{rfc_8879}, allowing endpoints to compress certificate chains using algorithms like zlib. However, the compression of Post-Quantum certificates is not well discussed as of now. Other methods include enabling hardware acceleration for TCP Segmentation Offload (TSO) and Large Receive Offload (LRO).

\textbf{2) Larger Key Exchanges: }

In addition to handshake overhead, post-quantum cryptography introduces rekeying challenges. TLS/DTLS-secured channels in 5G networks require frequent rekeying to maintain forward secrecy and mitigate nonce reuse vulnerabilities inherent in GCM modes~\cite{aes_gcm_issue}. Post-Quantum KEMs increase the computational and bandwidth costs of rekeying, contributing to an additional transport-layer stress.

Even in non–resource-constrained 5G infrastructure, these issues can manifest as:
\begin{itemize}
    \item Increased handshake and rekeying latency across control and data-plane interfaces.
    \item Greater susceptibility to MTU fragmentation and packet loss, especially over DTLS.
    \item Suboptimal TCP performance due to limitations in window scaling during large initial handshakes.
    \item Delays in control-plane responsiveness during orchestration, scaling, or mobility events.
    \item Incompatibilities with middleboxes or firewalls enforcing strict packet size or fragmentation policies.
\end{itemize}

\textbf{SCTP Encapsulation by PQ-DTLS 1.3}: As specified in~\cite{rfc_8261}, when SCTP packets are encapsulated by DTLS, the protocol must respect SCTP’s inherent packet reordering and reassembly mechanisms and avoid using compression algorithms. Large post-quantum (PQ) handshake messages can pose challenges for SCTP, as Path MTU Discovery (PMTUD) may not fully accommodate the increased sizes of keys and certificates. This can result in fragmentation at the IP or SCTP layers, introducing additional latency and processing overhead on both communicating peers. IP fragmentation is generally discouraged and considered fragile \cite{rfc_8900}.

To ensure reliable and efficient integration of post-quantum TLS in 5G networks,  requires several mitigation strategies. Techniques like certificate compression~\cite{rfc_8879}, tuning of TCP window sizes, DTLS 1.3 fragmentation~\cite{rfc_9147}, and adjustments to MTUs and middlebox behavior are all needed to manage the larger payloads introduced by PQC. Interoperability remains a key concern, especially since many vendors—particularly those supplying Physical Network Functions (PNFs)—have not yet adopted PQ algorithms or protocol extensions. This is particularly noticeable on radio-facing links like the O-FH M-Plane, where older equipment typically lacks support for PQ-enabled control and management.

\subsection{Standardization Challenges}

\begin{itemize}
  \item \textbf{Multi-Vendor Environment:} O-RAN’s disaggregated RAN architecture comprises components from multiple vendors, which increases the challenge of ensuring interoperability and consistency across PQC implementations.

\item \textbf{Diverse Network Architecture:} O-RAN defines numerous interfaces (e.g., F1, N2, N3), each with distinct security and transport requirements. Integrating post-quantum cryptography for each interface increases complexity and necessitates customized testing and development.

    \item \textbf{Lack of Unified Conformance tests:} The lack of conformance tests for stability, connectivity in vendor-neutral O-RAN deployments adds hindrance to stable deployments of PQC-enabled RAN.

    \item \textbf{Developing Standards:} Post-quantum cryptography standards continue to evolve, with many algorithms (e.g., ML-KEM, ML-DSA) only recently standardizePost-quantum cryptography standards continue to evolved by NIST. PQ alternatives of protocols are still under development and performance testing. This uncertainty and fast-moving nature make it difficult to define stable, long-term cryptographic baselines for O-RAN interfaces, which delays integration into official O-RAN Alliance specifications.
    \item \textbf{Limited Expertise in PQC:} A shortage of engineers skilled in both O-RAN and PQC has slowed down practical adoption and standardization efforts.

\end{itemize}

\section{Q-RAN: Implementation Details}

In this section we provide detailed implementation specifications for Q-RAN, the quantum-secure O-RAN solution. We provide protocol specifications, software libraries, and real solution logs from testing and evaluation environments.

\begin{figure*}[t]
    \centering
    \includegraphics[width=0.9\textwidth]{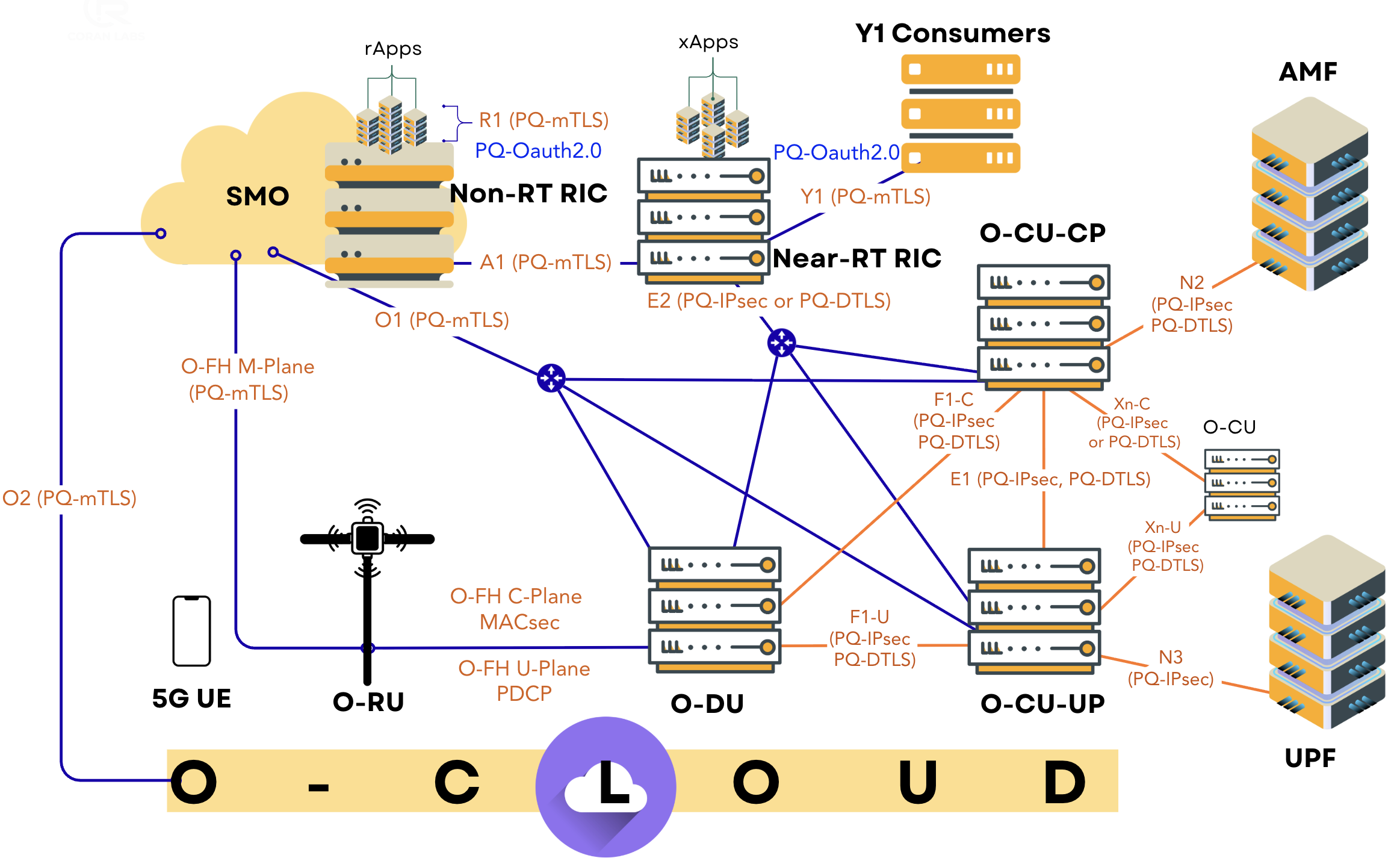}
    \caption{Q-RAN Security Architecture showing complete integration of PQ-mTLS, PQ-IPsec/PQ-DTLS, PQ-OAuth2.0 across all O-RAN interfaces connecting SMO, Non-RT RIC, Near-RT RIC, O-CU, O-DU, O-RU, O-Cloud, 5G UE, AMF, and UPF.}
    \label{fig:qran}
\end{figure*}

Figure~\ref{fig:qran} illustrates the complete Q-RAN security architecture.

\subsection{PQ-TLS/DTLS Enhancement Specifications}

The Q-RAN solution integrates comprehensive, well-tested quantum-resistant upgrades into TLS and DTLS, ensuring cryptographic robustness across the entire lifecycle—from key generation and exchange to authentication and encryption. Building on the protocol-level upgrades outlined earlier, we now summarize the key architectural and cryptographic design choices made to implement a robust, forward-compatible security framework. These include:

\subsubsection{Hybrid Key Exchange Mechanisms}

Q-RAN supports both pure and hybrid PQ key exchange methods. The most commonly used hybrid method is summarized below:

\textbf{X-Wing:} A hybrid KEM, X-Wing combines X25519 and ML-KEM-768 into a general-purpose KEM designed for efficiency and simplicity, as detailed in \cite{draft_xwing}. This construction is suitable for use in protocols like TLS and DTLS. X-Wing is IND-CCA secure, with overall security bounded by the respective strengths of ML-KEM-768 and Curve25519 (gap C-DH). The hybrid KEM remains secure under the assumption that SHA-3 is collision- and preimage-resistant. X-Wing achieves superior performance by introducing optimizations such as omitting the hashing of ML-KEM ciphertexts and flattening DHKEM, a feature often missing in other hybrid KEMs. It also extends plain ML-KEM by providing additional binding properties, including MAL-BIND-K, CT-PK, and LEAK-BIND-K-CT. As described in \cite{draft_xwing}, the key sizes for this hybrid KEM are 1184 octets for the public key and 2464 octets for the private key, with a ciphertext size of 1120 octets.

\subsubsection{Hybrid Digital Signature Algorithms}

Quantum-resistant authentication requires upgrading digital signatures using composite approaches that combine classical and post-quantum algorithms. A signature is valid only if both components verify correctly.

The hybrid signature scheme adopted in Q-RAN is \textbf{Ed448-ML-DSA-65}, which guarantees SUF-CMA (Strong Unforgeability under Chosen Message Attack) due to the individual properties of its underlying schemes \cite{draft_lamps_certs}. The composite scheme utilizes pre-hashing for the Ed448 signature using the XOF SHAKE256. Pre-hashing is deliberately avoided for ML-DSA to mitigate potential collision risks \cite{draft_lamps_certs}. We further incorporate context strings and algorithm-specific prefixes into the signing process and disallow standalone use of individual keys to achieve stronger non-separability (SNS), thereby preventing an attacker from producing a valid signature using only one component. \footnote{Note: All modes of ML-DSA and ML-KEM are supported in the Q-RAN solution, along with both pure PQ and hybrid configurations, providing flexibility for different security and operational requirements.}

\subsection{Quantum-Resistant DTLS Implementation}

\begin{figure*}[t]
    \centering
    \includegraphics[width=0.8\linewidth]{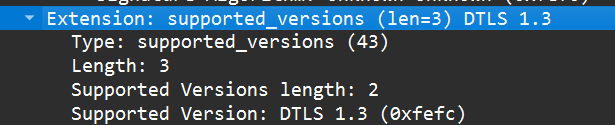}
    \caption{Packet Capture for DTLS 1.3}
    \label{fig:pcap-dtls1.3}
\end{figure*}

DTLS 1.3 was selected for Q-RAN as it is the only version compatible with the addition of post-quantum key exchanges and signatures. DTLS 1.3 also provides mandatory forward secrecy and robust fragmentation—improvements over the optional forward secrecy and brittle fragmentation mechanisms in DTLS 1.2. The inclusion of frequent rekeying and reduced round-trip times (RTT) in version 1.3 enhances both security and performance. Figure~\ref{fig:pcap-dtls1.3} shows the packet capture logs for DTLS 1.3.

The post-quantum hybrid suites described earlier are reused in DTLS 1.3, wherein the client initiates the key share and the server responds with the PQ-KEM ciphertext. PQ certificate chains and signatures are employed in the Certificate and CertificateVerify payloads. Q-RAN also supports experimental testing with other PQ KEMs (HQC, BIKE, NTRU, Saber, FrodoKEM) and signature schemes (SLH, Falcon, CROSS, LMS/XMSS).
\\

\begin{figure*}[t]
\centering
\hspace{-1.2cm}
\includegraphics[width=0.55\textwidth]{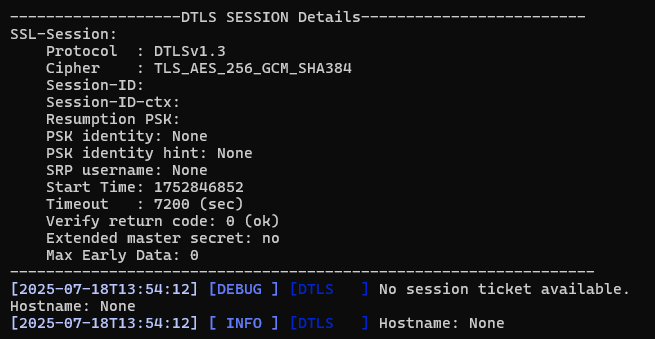}
\caption{PQ DTLS 1.3 Session between CU and DU}
\label{fig:dtls-sess}
\end{figure*}

A practical demonstration of Post-Quantum DTLS 1.3 on the midhaul interface is available in the video \href{https://www.youtube.com/watch?v=rref2-2oyU8}{\textit{Securing Midhaul Interface Using PQ-DTLS 1.3}}. The Post-Quantum DTLS 1.3 session details are briefed in Figure~\ref{fig:dtls-sess}.

\begin{figure*}
\centering
\includegraphics[width=0.8\textwidth]{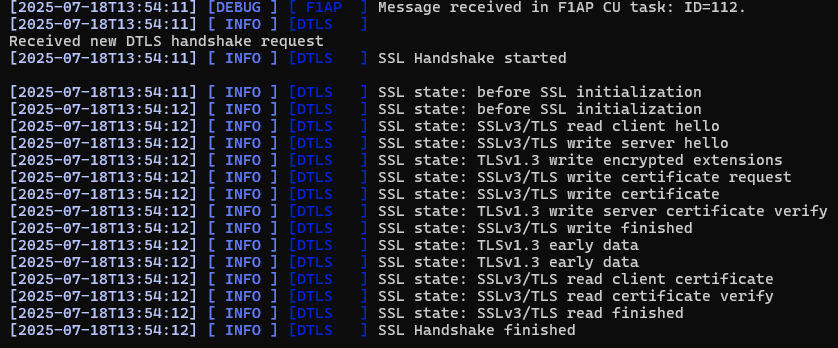}
\caption{SSL States at each stage of the handshake}
\label{fig:dtls-handshake}
\end{figure*}

\begin{figure*}[t]
\centering
\includegraphics[width=\textwidth]{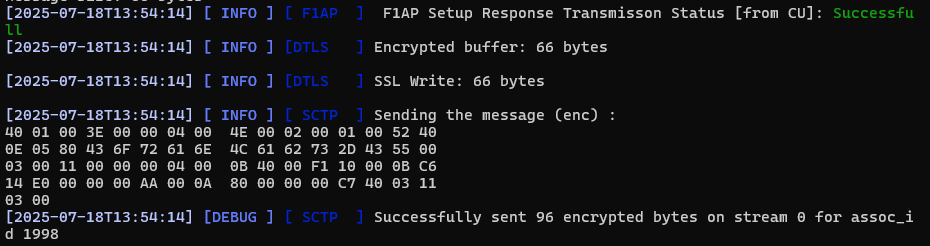}
\caption{F1AP Messages under DTLS encryption}
\label{fig:dtls-hs-enc}
\end{figure*}

\subsubsection{Certificate-Based Authentication}

Authentication is anchored by a Post-Quantum Certificate Authority (PQ-CA) logically hosted within the SMO framework. The SMO issues and manages quantum-resistant identities for all network components, including RICs, CUs, DUs, and applications.

The PQ-CA issues X.509 certificates using composite signature algorithms. The \texttt{p384mldsa65} specification represents a hybrid signature combining:
\begin{itemize}
\item \textbf{p384:} Classical ECDSA using the NIST P-384 curve, a widely trusted standard.
\item \textbf{mldsa65:} NIST-standardized post-quantum ML-DSA using the \texttt{ml-dsa-65} parameter set, offering an excellent security-performance balance.
\end{itemize}

This composite approach, aligned with current IETF drafts, ensures signatures are valid only when verified by both classical ECDSA and post-quantum ML-DSA algorithms, providing crypto-agility and backward compatibility.

\subsubsection{Handshake and Symmetric Encryption}

While authentication uses composite signatures, DTLS 1.3 handshakes are hardened through hybrid KEMs that combine classical key agreement (ECDHE) with PQC KEMs (ML-KEM). Figure~\ref{fig:dtls-handshake} shows the various Handshake states for DTLS 1.3 during the handshake.

Once the hybrid handshake completes and shared secrets are established, the connection transitions to symmetric encryption for subsequent data transfer. The specified cipher suite, \texttt{TLS\_AES\_256\_GCM\_SHA384}, provides strong, standardized encryption for TLS 1.3:
\begin{itemize}
\item \textbf{AES-256-GCM:} AES-256 is widely considered quantum-resistant, making it the preferred choice for symmetric encryption. GCM mode provides AEAD (Authenticated Encryption with Associated Data), ensuring both data integrity and confidentiality via GHASH and AES-CTR operations. Due to AES-GCM’s nonce reuse restrictions, extended nonce variants (e.g., XAES-GCM) may be used for high-throughput RAN interfaces.
\item \textbf{SHA-384:} The SHA-384 hash function (SHA-2 family) is employed in the TLS/DTLS key schedule, specifically in the HKDF-based key derivation process (\texttt{Extract-Expand}), to produce pseudorandom secrets and expand them into subkeys (e.g., MAC keys, AEAD keys).
\end{itemize}

Given the high throughput of 5G interfaces, it frequently triggers DTLS 1.3 key updates to maintain forward secrecy and limit key reuse. Figure~\ref{fig:dtls-hs-enc} depicts the encryption of F1AP messages at the DU.
The implementation (with support for split 7.2) is available at \textbf{\href{https://github.com/coranlabs/Q-RAN}{github.com/coranlabs/Q-RAN}}.

\subsubsection{QRNG Integration}

All PQC algorithms rely critically on high-entropy random numbers. Q-RAN integrates Quantum Random Number Generators (QRNGs) as entropy sources for all cryptographic operations.

\textbf{Ephemeral Key Generation:} For each DTLS session, hybrid KEMs require ephemeral key pairs. QRNGs supply high-entropy seeds to ensure these keys are unpredictable and independent across sessions.

\textbf{Certificate and Signature Generation:} The PQ-CA within the SMO uses QRNGs to generate long-term private keys for composite certificates. Additionally, ML-DSA depends on QRNGs for nonce generation, ensuring each signature operation is unique and resistant to private-key leakage attacks.

\subsection{Quantum-Resistant mTLS Implementation}

\begin{figure*}
\centering
\includegraphics[width=0.8\linewidth]{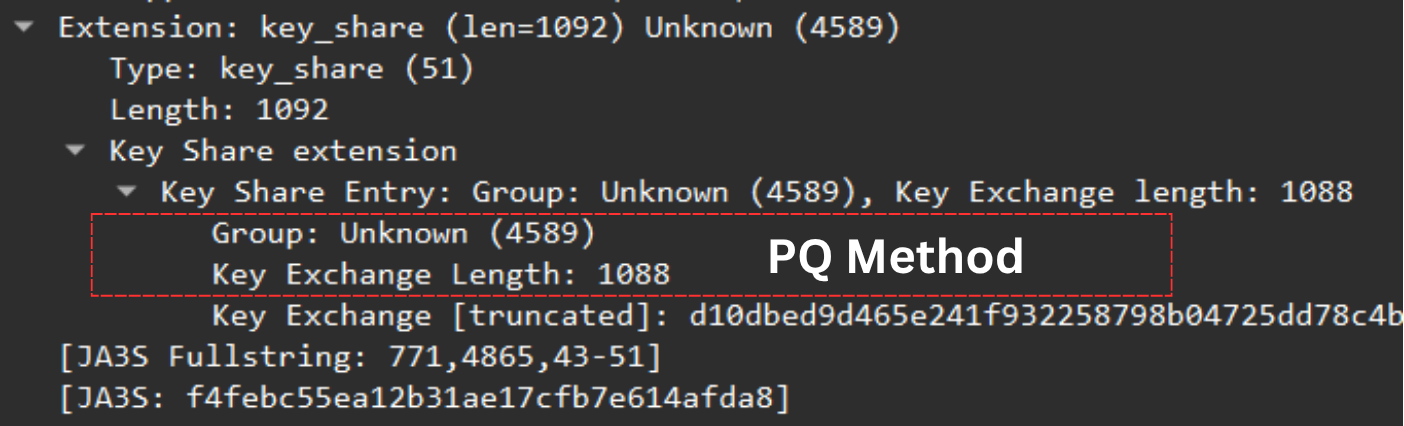}
\caption{Packet capture for Post-Quantum KEMs in TLS 1.3}
\label{fig:mlkem_tls_pcap}
\end{figure*}

Following O-RAN specifications \cite{oran_security}, Q-RAN adopts TLS 1.3 with post-quantum extensions similar to the PQ-DTLS 1.3 design described above. The same primitives are reused to maintain uniform security and cryptographic consistency across the RAN. We also acknowledge ongoing IETF developments such as \cite{draft_tls_ecdhe_mlkem}, \cite{composite_mldsa_draft}, and \cite{mldsa_tls_draft}.
Figure~\ref{fig:mlkem_tls_pcap} shows ML-KEM as the key exchange in the ClientHello message of TLS 1.3.

\subsubsection{Authentication Architecture}

The private PQ-CA hosted within the SMO issues quantum-resistant X.509 certificates to all O-RAN components, establishing a verifiable quantum-safe root of trust.
Certificates employ hybrid signature schemes combining classical and post-quantum algorithms—typically NIST-standardized ML-DSA (e.g., ML-DSA-65) with Ed448 or Ed25519. Signatures are valid only if both the classical and quantum components verify successfully, ensuring resilience even if one algorithm is compromised.

\subsection{Quantum-Resistant IPsec Implementation}

\begin{figure*}[t]
    \centering
    \includegraphics[width=0.8\linewidth]{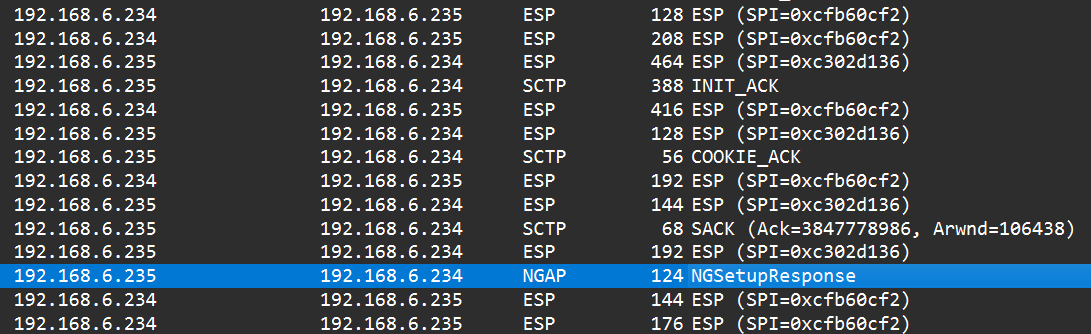}
    \caption{Packet Capture for IPsec ESP encryption and decryption for SCTP Packets}
    \label{fig:esp-sctp}
\end{figure*}

Q-RAN's IPsec stack employs a multi-layered quantum-resistant approach built upon the strongSwan IKEv2 implementation. strongSwan’s maturity and extensibility enable integration of modern post-quantum extensions. Packet capture for IPsec ESP working in conjunction with NGAP (SCTP) shown in Figure~\ref{fig:esp-sctp}.

\subsubsection{Hybrid Key Exchange and Authentication Implementation}

% The core mechanism is a hybrid key exchange that combines classical algorithms with PQC KEMs, ensuring crypto-agility—connections remain secure as long as at least one algorithm remains unbroken.

% The selected PQC algorithm, ML-KEM-768 (FIPS 203), provides AES-192–equivalent security and broad implementation support. Key exchange uses extensions defined in RFCs 9242 and 9370, supported in strongSwan version 6.0.0 and later.

% Initial \texttt{IKE\_SA\_INIT} exchanges establish a secure channel using classical algorithms (e.g., ECDH with Curve25519). Subsequently, ML-KEM-768 key exchanges occur within encrypted \texttt{IKE\_INTERMEDIATE} messages—a necessary design choice to handle large PQC key sizes that cannot be fragmented in initial handshakes. In strongSwan, this hybrid flow is configured using syntax such as \texttt{ke1\_mlkem768}.
\renewcommand{\arraystretch}{1.3}
\small
\setlength{\tabcolsep}{6pt}
\begin{table*}[!h]
\centering
\begin{tabular}{>{\raggedright}p{3cm} p{10cm} >{\raggedright\arraybackslash}p{3cm}}
\toprule
\textbf{Direction} & \textbf{Message Content} & \textbf{Post-Quantum Component} \\ 
\midrule
\rowcolor{lightgray}
\textbf{\texttt{IKE\_SA\_INIT}}\\ Initiator $\rightarrow$ Responder & 
HDR, SAi1 (AES-CBC-256, HMAC-SHA2-256, PRF-HMAC-SHA2-256, DH14), KEi, Ni, N(INTERMEDIATE\_EXCHANGE\_SUPPORTED), [CERT] &
-- \\

Responder $\rightarrow$ Initiator &
HDR, SAr1, KEr, Nr, [CERTREQ], N(INTERMEDIATE\_EXCHANGE\_SUPPORTED) &
-- \\
\midrule
\rowcolor{lightgray}
\textbf{\texttt{IKE\_INTERMEDIATE}}\\Initiator $\rightarrow$ Responder &
HDR, SK\{ KEi(n) (\textbf{ML-KEM-768 public key}) \} &
\textbf{ML-KEM-768} \\

Responder $\rightarrow$ Initiator &
HDR, SK\{ KEr(n) (\textbf{ML-KEM-768 public key}) \} &
\textbf{ML-KEM-768} \\
\midrule
\rowcolor{lightgray}
\textbf{\texttt{IKE\_AUTH}}\\Initiator $\rightarrow$ Responder &
HDR, SK\{ IDi, [CERT], [IDr], AUTH (with \textbf{PPK}), SAi2 (ESP proposal), TSi, TSr, N(\textbf{PPK\_IDENTITY, PPK\_ID}) \} &
\textbf{PPK} \\

Responder $\rightarrow$ Initiator &
HDR, SK\{ IDr, [CERT], AUTH, SAr2, TSi, TSr, N(\textbf{PPK\_IDENTITY}) \} &
\textbf{PPK} \\
\midrule
\rowcolor{lightgray}
\textbf{\texttt{CREATE\_CHILD\_SA}}\\Initiator $\rightarrow$ Responder &
HDR, SK\{ SA (ESP proposal), Ni, [KEi], TSi, TSr \} &
(Optional \textbf{PQ KEM}) \\

Responder $\rightarrow$ Initiator &
HDR, SK\{ SA (selected), Nr, [KEr], TSi, TSr \} &
(Optional \textbf{PQ KEM}) \\
\midrule
\rowcolor{lightgray}
\textbf{\texttt{INFORMATIONAL}}\\Initiator $\rightarrow$ Responder &
HDR, SK\{ DELETE, NOTIFY: NAT\_DETECTION\_SOURCE\_IP \} &
-- \\

Responder $\rightarrow$ Initiator &
HDR, SK\{ DELETE Response / ACK \} &
-- \\
\bottomrule
\end{tabular}
\caption{IPsec (IKEv2) message flow with post-quantum components highlighted.}
\label{tab:ipsec-pq-flow}
\end{table*}

\vspace{1em}
\noindent\textit{Notes (Table \ref{tab:ipsec-pq-flow}:} 
\begin{itemize}
\item \textbf{ML-KEM-768} provides post-quantum key encapsulation (RFC 9370).
\item \textbf{PPKs} offer hybrid PQ authentication (RFC 8784) — 256-bit classical, $\approx$128-bit quantum-safe.
\item PQ KEM to strengthen rekeying (CREATE\_CHILD\_SA).
\end{itemize}

\subsubsection{strongSwan Library Configuration}

This quantum-resistant IPsec solution is implemented using strongSwan library, which can be compiled with Open Quantum Safe (OQS) liboqs plugin to enable support for necessary PQC algorithms. Latest strongSwan versions (6.0.0$+$) have begun integrating this functionality more directly, streamlining configuration of hybrid key exchanges.

Example strongSwan configuration for PQ-IPsec:
\begin{verbatim}
{conn pq-ikev2
    ppk=yes
    ppk_id=qrng-generated-ppk-id
child pq-ipsec-tunnel{
    ike=aes256-sha384-ecp384-ke1_mlkem768!
    esp=aes256gcm128-sha384-ecp384!
    }
}
\end{verbatim}

\subsubsection{Entropy Source Integration}

The high-quality randomness provided by QRNGs, as discussed previously, is securely provisioned to each RAN component through encrypted peripheral links, preventing eavesdropping or tampering. The seeds and the required purpose are described in the Table~\ref{tab:seeds-alg}.

\begin{table}[h!]
\centering
\footnotesize
% \hspace{-1cm}
\setlength{\tabcolsep}{3pt}
\renewcommand{\arraystretch}{1.4}
\begin{tabular}{p{2.7cm} p{1.8cm} p{1.8cm} p{1.4cm}}
\rowcolor{lightblue}
\textbf{Scheme} &
\textbf{Keygen Seed} &
\textbf{Encap Seed} &
\textbf{Sign Seed} \\
\hline
ML-KEM-768 & 64 B & 32 B & — \\
ML-DSA-65 & 32 B & — & 32 B (opt.) \\
X-Wing (Hybrid) & 32 B & 32 B & — \\
Ed448-ML-DSA-65 & 57 B & — & 32 B (opt.) \\
\end{tabular}
\caption{Seed sizes (bytes) used in key generation, encapsulation, and signing for various schemes.}
\label{tab:seeds-alg}
\end{table}

% This entropy can also be combined with kernel-level CSPRNG/TRNG sources, as shown in Figure~\ref{fig:qrng}.

% \begin{figure*}
% \centering
% \includegraphics[width=0.8\linewidth]{figures/qrng.png}
% \caption{Accumulating entropy from QRNG and DRBG sources}
% \label{fig:qrng}
% \end{figure*}

% \begin{figure*}
%     \centering
%     \includegraphics[width=0.8\linewidth]{figures/x25519kyber.png}
%     \caption{Packet Capture for X25519Kyber768 in TLS 1.3}
%     \label{fig:kyber_tls_pcap}
% \end{figure*}

% \subsubsection{Post-Quantum Pre-shared Keys}

% The solution incorporates Post-Quantum Pre-shared Keys (PPKs), also termed PQ-PSKs, as specified in RFC 8784, providing additional quantum resistance layers.

% High-entropy, quantum-generated pre-shared keys guaranteeing 256 bits of classical security are provisioned out-of-band to communicating peers. PPKs are not exchanged during handshakes but are mixed into key derivation functions. This means even if adversaries record entire IKEv2 exchanges and later use quantum computers to break asymmetric cryptography, they still cannot derive final session keys without PPKs, rendering harvested data useless.

% PPK security is entirely dependent on unpredictability. Therefore, PPKs must be generated using QRNGs to ensure highest possible entropy, making them truly random secrets immune to prediction. strongSwan supports PPKs as defined in RFC 8784.

\subsubsection{Certificate-Based Authentication}

Digital certificates used for IKEv2 peer authentication must be quantum-resistant. These certificates are issued by centralized Post-Quantum Certificate Authority (PQ-CA) located on the SMO. The PQ-CA uses composite signature schemes like p384mldsa65 to sign certificates, ensuring they are trusted and verifiable in both classical and post-quantum environments. The composite signature schemes are thought to provide dual-security until active quantum adversaries are not present, and provide an easier transition to pure Post-Quantum methods.

QRNGs serve as roots of trust for all randomness, used to generate high-entropy seeds for all ephemeral keys in Asymmetric Key exchanges and for PPKs. Furthermore, private keys for certificates issued by PQ-CA are also generated using QRNGs, ensuring entire cryptographic lifecycles and certificate truststore are secured with provably unpredictable randomness.

\subsection{Post-Quantum SMO Security Architecture}

\begin{figure*}
    \centering
    \includegraphics[width=0.75\linewidth]{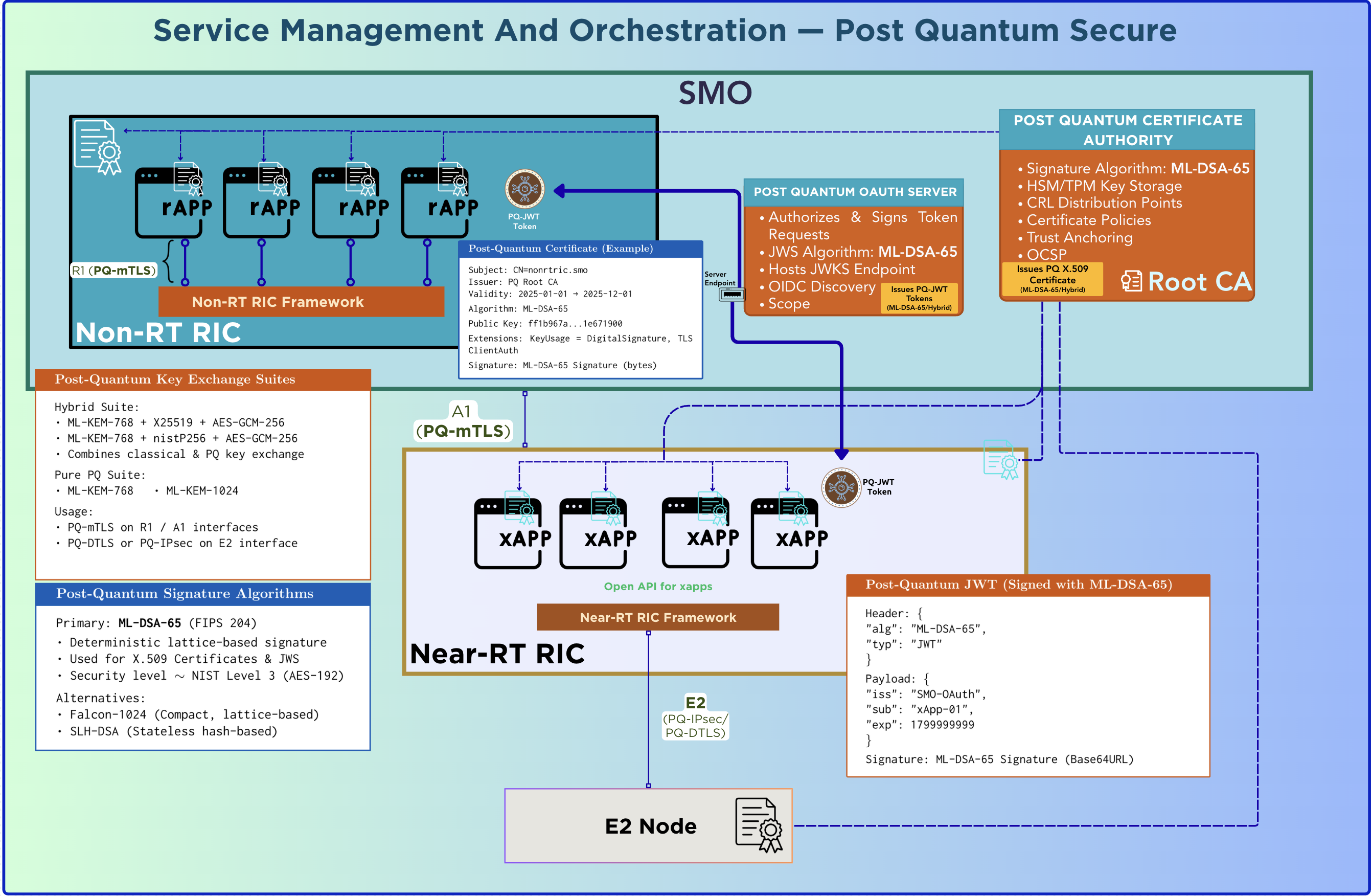}
    \caption{Post-Quantum Secure SMO architecture integrating PQ-mTLS, PQ-IPSec/PQ-DTLS, and PQ-OAuth 2.0 across Non-RT RIC, Near-RT RIC, and E2 nodes, enabling quantum-resilient authentication, authorization, and communication through ML-KEM key exchange and ML-DSA-65-based certificates and tokens.}
    \label{fig:pq-smo}
\end{figure*}

The SMO security strategy focuses on robustly protecting all communication channels—both external interfaces (O1, O2, A1, R1) and internal communications between microservices (SMOs) as shown in Figure~\ref{fig:pq-smo} . This protection is provided via PQ-mTLS, and PQ-OAuth 2.0 between the participating entities.

\subsubsection{PQ-mTLS 1.3 for All Communications}

The mandated mechanism is mutual Transport Layer Security (mTLS), specifically based on TLS 1.3, enhanced to incorporate post-quantum (PQ) cryptography. The upgrade process aligns closely with proposals such as PQ-(D)TLS 1.3, but includes modifications to the  \texttt{ClientCertificate} and  \texttt{ClientCertificateVerify} messages on client side to support post-quantum X.509 certificates and post-quantum digital signatures.% This PQ-mTLS implementation relies on two core PQC components

% \textbf{Post-Quantum Certificates:} All O-RAN functions, including SMO and elements it communicates with, are provisioned with X.509 certificates issued by central Post-Quantum Certificate Authority (PQ-CA) acting as root of trust. Certificates use PQC digital signature algorithms such as ML-DSA for signatures. Certificates contain composite public keys and are signed by PQ-CA with composite signatures combining classical algorithms (e.g., ECDSA using NIST p384 curve) with post-quantum ones (ML-DSA at NIST security level 3). Relying parties must validate both signatures for certificates to be considered authentic, providing hedges against potential weaknesses in either classical or new PQC algorithms.

% \textbf{Hybrid Key Exchange:} TLS 1.3 handshakes are modified to use hybrid key exchange mechanisms. Clients and servers exchange keying material for both classical key exchange algorithms (e.g., ECDHE with X25519) and PQC Key Encapsulation Mechanisms (e.g., ML-KEM-768). Final shared secrets derive from outputs of both algorithms, ensuring sessions remain secure as long as at least one component algorithm remains unbroken.

\subsubsection{Centralized Public Key Infrastructure}

Beyond securing its own communications, the SMO functions as central PKI governance engine for the entire RAN, acting as management and registration authority for underlying Post-Quantum Certificate Authority (PQ-CA). The PQ-CA is the core trust anchor responsible for signing all quantum-resistant identities, while the SMO orchestrates complex lifecycles of these PQC certificates.

This includes automated issuance, renewal, distribution, and revocation for every network function, effectively translating operator policy into actions performed by PQ-CA. Centralized management is essential for enforcing consistent cryptographic policies and achieving crypto-agility necessary to respond to future threats.

\subsubsection{Quantum-Resistant Application Authorization}

RAN Intelligent Controllers (Non-RT RIC and Near-RT RIC) and third-party applications they host (rApps and xApps) are O-RAN's innovation engines, enabling AI/ML-driven network optimization. However, allowing third-party code to control RAN introduces significant security risks. A robust, quantum-resistant authorization framework is required.

The O-RAN Alliance specifies OAuth 2.0 framework for this purpose. A quantum-resistant implementation, termed PQ-OAuth, provides granular, token-based authorization aligning with Zero Trust Architecture principles \cite{oran_zta} \cite{nist_zta}.

\paragraph{Authorization Architecture Roles}
\begin{figure*}
    \centering
    \includegraphics[width=0.95\linewidth]{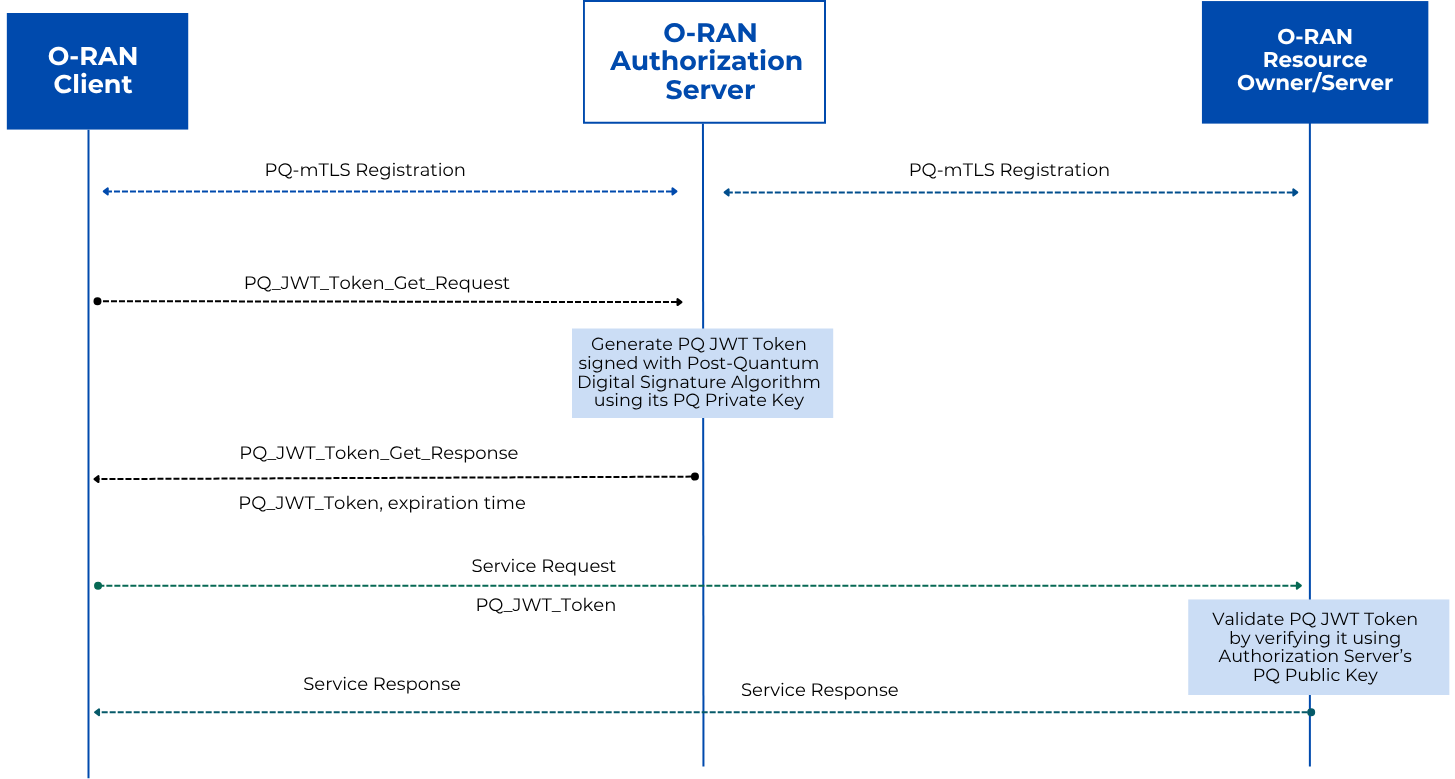}
    \caption{PQ-OAuth 2.0 Flow in Open RAN components}
    \label{fig:oauth-jwt-oran}
\end{figure*}

\begin{itemize}
    \item \textbf{Authorization Server(AS):} Logical role fulfilled by dedicated service within SMO framework, responsible for authenticating clients and issuing access tokens. The AS also hosts a JSON Web Keys Set (JWKS) \cite{rfc_7517} endpoint for clients to fetch the PQ verification keys in JSON format.
    \item \textbf{Client:} rApp or xApp requesting access to protected resources.
    \item \textbf{Resource Server:} O-RAN service or function exposing protected APIs (e.g., R1 interface service provider within SMO for rApps, A1 interface on Near-RT RIC for xApps)
    \item \textbf{Resource Owner:} Network operator defining authorization policies enforced by Authorization Server
\end{itemize}

\paragraph{Quantum-Resistant Authorization Flow}

\textbf{Onboarding and Authentication:} rApps/xApps are onboarded onto appropriate RIC platforms by SMO. During onboarding, SMO facilitates requests to Post-Quantum Certificate Authority (PQ-CA), which provisions applications with long-term PQ digital certificates (e.g., composite p384ml-dsa65 certificates).

\textbf{Token Request:} To access protected resources, rApps/xApps initiate PQ-mTLS connections to Authorization Server's token endpoints in SMO. They authenticate themselves using their provisioned PQ certificates, then send requests for access tokens specifying desired scopes (e.g., o-ran-smo:performance-data:read or o-ran-a1:policy:write).

\textbf{Token Issuance:} Authorization Servers verify rApp/xApp identities via mTLS handshakes, consult policy engines to determine if specific applications are authorized for requested scopes. If authorized, they generate JSON Web Tokens (JWT).

\textbf{PQ-Signed JWT:} JWT security is paramount. Authorization Servers sign JWTs using high-assurance private keys. The ed448-dilithium3 specification points to composite signature schemes where JWT headers and payloads are signed separately by both Ed448 (classical) and ML-DSA (Dilithium at level 3) algorithms, with resulting signatures concatenated to form final JWT signatures. JWT header alg parameters contain unique identifiers registered for specific composite schemes, ensuring token integrity and authenticity are protected against both classical and quantum adversaries---active or passive.

\textbf{Resource Access:} rApps/xApps receive signed JWTs, then make requests to Resource Servers (e.g., API calls over R1 interfaces), including JWTs in Authorization HTTP headers as Bearer tokens.

\textbf{Token Validation:} Resource Servers receive requests and must validate JWTs before processing. They retrieve Authorization Server public keys. Trust in these keys is established because Authorization Server identity certificates were issued by network root PQ-CA, which Resource Servers are configured to trust. Resource Servers use trusted composite public keys to verify both Ed448 and ML-DSA signatures on tokens, also validating token claims such as expiration times (exp), intended audiences (aud), and ensuring requested actions are permitted by token scope claims. Only if all checks pass are requests processed.

The token authentication flow described above is summarized in the Figure~\ref{fig:oauth-jwt-oran}.

\noindent
\\A demonstration of the described implementation of Q-RAN, also featuring the usage of our Post-Quantum enabled 5G Core (QORE) and PKI solution is present at \textbf{\href{https://www.youtube.com/watch?v=yiH2O24eUWk}{YouTube}}.

\section{Future Works}

This section outlines the prospective directions and development opportunities for advancing quantum-resistant Open RAN (Q-RAN) security and interoperability. Future work focuses on expanding the scope of post-quantum cryptographic (PQC) frameworks across different layers of the 5G Network, enhancing interoperability through xFAPI, integrating PQC with open-source RAN stacks, alongside leveraging hardware acceleration, and extending protection to radio and management interfaces.

\subsection{xFAPI-PQC Integration for Vendor-Neutral Interoperability}

\begin{figure*}[!h]
    \centering
    \includegraphics[width=1.1\textwidth]{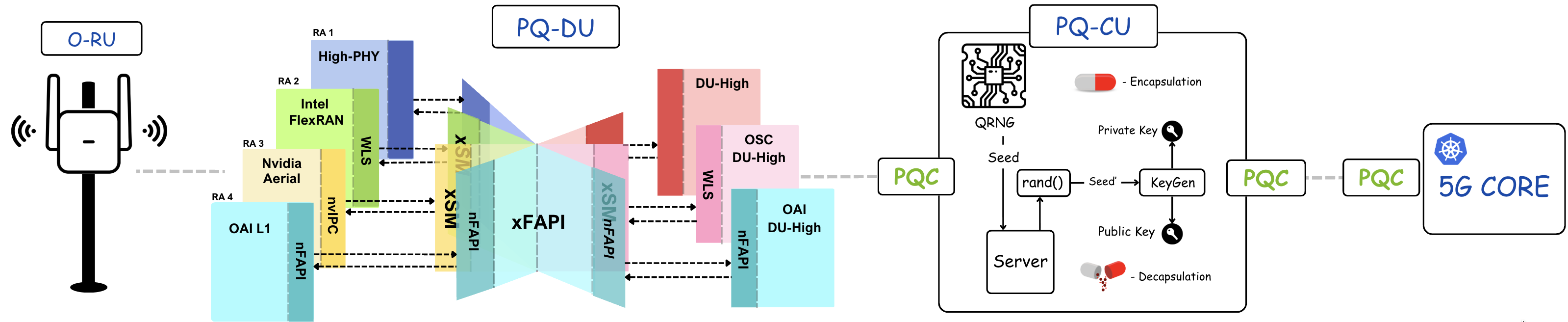}
    \caption{PQC-Enhanced xFAPI Architecture — illustrating secure interoperability between multi-vendor RUs and DUs through PQ extensions for SCTP and MAC Layers}
    \label{fig:xfapi_stack}
\end{figure*}

xFAPI \cite{xfapi_crl} is an interoperability framework designed to eliminate vendor lock-in within the Open RAN ecosystem by providing a standardized mechanism to connect any Radio Unit (RU) to any Distributed Unit (DU). It achieves this by employing inter-process communication (IPC) calls and SCTP-based data transfer mechanisms to ensure efficient interaction between disaggregated components.

To protect xFAPI against quantum-era attacks, we propose adding post-quantum cryptography either at the MAC layer or by encapsulating / layering SCTP traffic using PQ-DTLS 1.3. These upgrades allow secure, low-latency communication between multi-vendor components, with cryptographic primitives tailored specifically for the FAPI's real-time signaling needs.

This design supports cryptographic agility while maintaining interoperability across vendors. As shown in Figure~\ref{fig:xfapi_stack}, the proposed PQC-enhanced xFAPI architecture is built to serve as a security foundation for future quantum-resilient RAN deployments.

\subsection{Extension to OSC and Other Open-Source RAN Platforms}

Future work will also focus on bringing post-quantum cryptography to the O-RAN Software Community (OSC) \cite{oran_sc} and other open-source RAN projects to expand the quantum-resistant safety net. The goal is to build standardized PQC implementations across open ecosystems, thereby encouraging interoperability, transparency, and long-term security against threats posed by quantum computers. Our presentation at O-RAN Working Group 11 \cite{crl_oran_wg} highlights the importance of introducing quantum-safe network architecture into Open RAN and provides insights on how this can be achieved smoothly.

\subsection{Hardware Acceleration Dedicated for PQC and QKD}

% To meet the performance and latency requirements of telecom networks, dedicated hardware acceleration for PQC operations will be essential. Future deployments can leverage programmable cryptographic accelerators (e.g., FPGA-based or ASIC implementations) to offload computationally intensive PQC operations, thereby maintaining real-time processing capability.
5G RAN is a time-sensitive, performance-driven network designed to serve users and other RAN components with low latency and high throughput. To achieve this, operators often rely on hardware accelerators such as FPGAs or ASICs. These cards handle compute-intensive tasks like error correction, DSP, cryptographic processing, and routing. However, current cryptographic accelerators lack support for PQC algorithms. To maintain network performance, hardware-specific PQC implementations will be needed, as RAN connections are long-lived and depend heavily on asymmetric cryptography.

In parallel, integrating Quantum Key Distribution (QKD) \cite{etsi_qkd} systems adds an extra layer of information-theoretic protection, offering provably unbreakable security for network and critical interfaces. QKD can distribute Pre-Shared Keys (such as PPKs, discussed earlier) out-of-band between communication entities, enabling symmetric-only, air-gapped networks and reducing reliance on PQC for key exchange.

\subsection{Extension to Radio Interfaces, MACsec, and M-Plane}

Early post-quantum migration efforts focused mainly on key exchange to mitigate HNDL attacks. As quantum computing advances, authentication mechanisms must also evolve to defend against “Trust Now, Forge Later” threats. Accordingly, upgrades have been introduced at the control plane and other key layers of the RAN stack. Future work will extend PQC to MACsec for Ethernet-based fronthaul links and to M-Plane communications between RUs and SMO/O-DU components through PQ-mTLS and PQ-SSH, while exploring methods to implement these enhancements at the RU level.

These extensions aim to deliver comprehensive, end-to-end quantum resilience across both wired and wireless interfaces, while remaining aligned with existing IEEE authentication/confidentiality and O-RAN security standards.

\subsection{PQ Blockchain RAN (B-RAN)}

Inspired by recent work on quantum-secure blockchain architectures~\cite{Thales2025QuantumResilient} in 5G RAN, we see a clear path to extending Q-RAN into a decentralized, blockchain-backed RAN system—B-RAN. This would allow RAN components and IoT devices to securely authenticate and exchange data using a distributed trust model. Post-Quantum secure identities could be anchored to the blockchain, and smart contracts could automate key operations like credential rotation, device revocation, or audit logging. As more edge and IoT devices start supporting post-quantum cryptography, it become necessary to secure these endpoints against future threats.

\section{Conclusion}

This paper introduced Q-RAN, a practical framework for securing Open RAN architectures against emerging quantum computing threats by systematically integrating post-quantum cryptography (PQC). We discussed the challenges of replacing quantum-vulnerable primitives in key telecommunications security protocols—including mTLS, DTLS, IPsec, and OAuth 2.0—with NIST-standardized ML-KEM and ML-DSA algorithms. The proposed hybrid PQC model provides immediate protection against both active and passive quantum adversaries while supporting a smoother transition from classical systems.

Our analysis shows that lattice-based PQC algorithms can meet the latency and throughput demands of disaggregated 5G infrastructures, and hardware acceleration and optimized implementations would greatly help in minimizing computational overhead.

Q-RAN mitigates the “Harvest Now, Decrypt Later” threat by enabling quantum-safe encryption and authentication across O-RAN interfaces—from fronthaul and midhaul to management and control planes. The framework integrates quantum-generated entropy through QRNGs and designates the Service Management and Orchestration (SMO) layer as a unified trust anchor, hosting the Post-Quantum Certificate Authority (PQ-CA) and PQ-OAuth 2.0 service to ensure end-to-end confidentiality and integrity across certificate chains.

Rolling out Q-RAN in real deployments will require tight collaboration between network operators, equipment vendors, and standards bodies like the O-RAN Alliance, 3GPP SA3~\cite{3gpp_sa3}, and the IETF. Standardization gaps, inconsistent PQ support across stacks, and integration at the edge (such as at RUs) are still open problems. Extending PQC across all relevant interfaces—without breaking existing workflows—remains a key focus.

The migration roadmap we've outlined lays out a practical path forward. With quantum computing evolving fast, it's important to start adapting networks now using proven solutions like Q-RAN—not just to stay ahead, but to keep critical systems secure and reliable.

\section*{Acknowledgments}

This work reflects the ongoing efforts of the \textbf{coRAN Labs} research team to advance quantum-safe technologies in next-generation telecom systems. We’d like to thank the \textbf{O-RAN Alliance} for shaping the open RAN landscape and driving key security specifications, and the \textbf{NIST} community for leading the standardization of post-quantum cryptography.

We also acknowledge the invaluable tools and support from the open-source ecosystem. The \textbf{Open Quantum Safe} project made it possible to experiment with real-world PQC stacks; the \textbf{strongSwan} team continues to push innovation in IKEv2 and IPsec; and the relevant \textbf{IETF} working groups have laid the groundwork for secure, post-quantum-ready Internet protocols. 

Finally, a big thanks to the \textbf{OpenAirInterface 5G} community—its open platform has been central to testing, integration, and validation throughout our project.

\bibliographystyle{IEEEtran} 
\bibliography{reference}

\end{document}